\documentclass[a4paper]{elsarticle}

\makeatletter

\usepackage{float}
\usepackage{latexsym}
\usepackage{graphicx}
\usepackage{amsmath} 
\usepackage{amsfonts} 
\usepackage{color}

\usepackage{amssymb,amsmath}
\usepackage{graphics,graphicx}
\graphicspath{{./figv10/}}

\DeclareGraphicsExtensions{.pdf,.eps}

\usepackage[english]{babel}
\usepackage{bbm,theorem}
\usepackage{a4wide}

{\theorembodyfont{\upshape}

}

\newcommand{\etal}{{\it\,et al.~}}

\usepackage[dvipsnames]{xcolor}

\title{Deep reinforced learning enables solving rich discrete-choice life cycle models to analyze social security reforms}

\begin{document} 
\selectlanguage{english}
\begin{frontmatter}

\author[EK]{Antti J. Tanskanen\corref{Corr}}
\ead{antti.tanskanen@ek.fi}
\address[EK]{Confederation of Finnish Industries EK, P.O. Box 30 (Etel\"{a}ranta 10), FI-00131 Helsinki, Finland}
\date{\today}

\begin{abstract}
Discrete-choice life cycle models of labor supply can be used to estimate how social security reforms influence employment rate. In a life cycle model, optimal employment choices during the life course of an individual must be solved. Mostly, life cycle models have been solved with dynamic programming, which is not feasible when the state space is large, as often is the case in a realistic life cycle model. Solving a complex life cycle model requires the use of approximate methods, such as reinforced learning algorithms. We compare how well a deep reinforced learning algorithm ACKTR and dynamic programming solve a relatively simple life cycle model. To analyze results, we use a selection of statistics and also compare the resulting optimal employment choices at various states. The statistics demonstrate that ACKTR yields almost as good results as dynamic programming. Qualitatively, dynamic programming yields more spiked aggregate employment profiles than ACKTR. The results obtained with ACKTR provide a good, yet not perfect, approximation to the results of dynamic programming. In addition to the baseline case, we analyze two social security reforms: (1) an increase of retirement age, and (2) universal basic income. Our results suggest that reinforced learning algorithms can be of significant value in developing social security reforms.
\end{abstract}
 
\end{frontmatter}

\section{Introduction}
Stochastic discrete-choice life cycle models of labor supply can be used to, e.g., estimate the impact of a social security reform on employment and unemployment rates (Hakola and Määttänen 2007, Määttänen 2014, Tanskanen 2020a).  Such models have also been used to analyze retiring (Jiménez‐Martín and Sánchez Martín 2007, Blau 2008) and internal return on pension premiums (Tanskanen, 2020b). Still, perhaps a more common issue analyzed with a life cycle model is optimal saving during the life cycle of an individual (Modigliani and Brumberg 1954).

In a stochastic life cycle model, actions and states are considered in the Markov decision process framework, that is, each decision is made based on the current state without considering the previous states, and the chosen action influences, but does not necessarily determine, the next state (Sutton and Barto 2018). Life cycle of each individual is followed for a relatively long time, e.g., for 50 years. The aim is to find for each state an action that maximizes the expected discounted future utility of an individual. Solving the life cycle model yields a recipe that tells which action to take at which state. This recipe can then be used to, e.g., deduce the aggregate employment rate. 

In a social security reform, the aim is often either to improve social security of certain individuals or to revise the benefits to encourage employment. Estimating how the reform affects behavior of individuals, and consequently the employment rate, is one of the things to consider when planning a social security reform. 
A change in employment can influence the financial sustainability of a social security scheme, which makes analysis of the implications of a social security reform important.
A typical microsimulation does not take behavioral changes into account, however, behavioral effects can, to a degree, be included using elasticities (Kotamäki\etal 2018). 

\subsection{Dynamic programming and reinforced learning}
Dynamic programming (DP) is a well-known method for solving a life cycle model (see, e.g., Heer and Maussner 2009,  Rust 1989, Määttänen 2014). It gives a good approximation to the optimal agent behavior, which converges towards the optimal solution in the limit of infinitely tight grid (e.g., Puterman 1990). However, applicability of dynamic programming is constrained by its large demand for resources and finite available computation time (Sutton and Barto 2018).

A realistic life cycle model describing a social security scheme often requires a large number of states. For example, it was estimated that to solve the life cycle model in Tanskanen (2020a), dynamic programming would need to solve the model for $10^{21}$ states. A large state space makes the use of dynamic programming unfeasible. Approximative methods are therefore needed (Rust 1997).  

Reinforced learning (often abbreviated RL) algorithms are approximative alternatives for dynamic programming in solving a life cycle model, in particular those with huge state spaces. Although theoretical knowledge of their convergence properties is not fully developed, their performance is good in practice, and it is supported by emerging theoretical results on the convergence (Wang\etal 2019). 

The main differences between dynamic programming and a deep reinforced learning algorithm are that a reinforced learning algorithm is typically based on sampling the life cycle trajectories (also known as episodes), and it does not depend on the model analyzed beyond the assumption that dynamics are described by a Markov decision process. When the state space is huge, approximate methods must also be able to generalize from the observed states, which is one of things deep reinforced learning methods are quite good at. 

\subsection{This study}
A reinforced learning algorithm often provides a good approximation to the optimal agent behavior. But is this approximation good enough to analyze how a social security reform influences optimal agent behavior? Resolving this question in the case of a relatively simple life cycle model is the aim of this study. We approach the issue in various ways: after describing the employed methods and the used life cycle model (Section 2), we compare the results obtained with dynamic programming and reinforced learning first as aggregates and statistics (Sections 3.1 and 3.2), and then the found action policies are compared in more detail (Section 3.3). We proceed on to present and compare the impact of two social security reforms: an increase of retirement age (Section 4.1) and implementation of universal basic income (Section 4.2). Finally, in Section 5 we discuss the obtained results and  provide some context in which to analyze the differences. The employed codes are freely available (Tanskanen, 2019a-c).

\section{Methods}
\subsection{Life cycle model}
The life cycle model considered here is a reduced version of the full life cycle model (Tanskanen, 2020a) aimed at capturing the Finnish social security scheme, and optimal labor supply in the presence of uncertainty in future wages. Here we only include three employment states in the life cycle model: employed, unemployed, and retired. Each agent receives automatically any social security to which the agent is entitle to, be it supplementary benefit, housing benefit, unemployment insurance (earnings-related or not), or retirement benefit. Net income of an agent is the sum of all benefits and wages minus income taxes and social security contributions. The life cycle model is largely fitted with data from Official Statistics of Finland (2020).

The utility (reward in the reinforced learning terminology) an agent receives during a time-step is measured as a logarithm of the agent's net income $n$ minus the value of free time lost $\kappa$, that is, 
\begin{equation}
u(n)=\log(n)-\kappa. 
\end{equation}
To an agent, a lower income from social security may be of the same or higher utility as working full-time due to the value of free time. Constant $\kappa=0.75$ representing the value of free-time is the same for all agents. This can be interpreted as stating that 1,800 euros per month net income working full-time has the same utility as 850 euros per month net income from social security, when there is no need to work. 
Equally, utility can be expressed as an equivalent net income taking into account the value of the lost free time. 

The life cycle of an individual is followed from the age of 18 until 70 years of age. Before age 70, mortality is not taken into account to keep the model simple. At the age of 70, the utility includes the discounted net value of future pensions and other benefits taking mortality into account. 

The time-step used in the simulation is one year. 
Each individual makes a decision to either stay in the current employment state or switch to another employment state at the beginning of each year. 
An employed person can either stay employed, or go unemployed. Similarly, an unemployed person can either stay unemployed or be employed. After the minimum retirement age, retirement is also available. 

\subsubsection{States}

The state space describing the life cycle model is 6-dimensional consisting of {\em employment state, age, accrued pension, the previous wage, time in the state,} and {\em the current wage}. 

State variable {\em time in the state} describes the time that has elapsed since the previous employment state transition. It is required to decide, whether an unemployed person is entitled to earnings-related unemployment benefit. 

Pension accrues at the rate of 1.5 percent of the wage each year, when the agent is employed. When the agent is unemployed, pension accrues at 1.125 percent of the previous wage for one year. After that, no earnings-related pension accrual is available. 

The wage process is stochastic. The next wage depends on the previous wage, the current wage, and a stochastic component.
Being unemployed reduces the next wage offer by 5 percent. The chosen action influences directly the employment state and indirectly all the other state variables except for age, which increases quite monotonically during the simulation. 

\subsubsection{Optimization problem}

Here we consider an agent interacting with a discounted Markov decision process
$(X , A, \gamma, P, u)$, as in Wu\etal (2017).
At each time $t$, an agent chooses the action that maximizes the present value of future utilities. In a policy gradient method, such as ACKTR, action $a_t\in A$ is chosen according to policy $\pi(a|S_t)$ given the current state $S_t\in X$. 
An action yields a utility $u(S_t, a_t)$ and a transition to the next state $S_{t+1}$ according to the transition probability $P(S_{t+1}|S_t, a_t)$. The agent aims at 
maximizing the expected $\gamma$-discounted cumulative return 
\begin{equation}
\label{eq:1}
J(\theta)=\max_{a\in A}E[\sum_{s=t}^T\gamma^su(n_s,S_s)],
\end{equation}
with respect to the policy parameters $\theta$.
Here $t$ is the analyzed time, $A$ is the set of available actions, $u(n,S)=(\log(n)+\kappa_S)/10$ is the scaled utility, $n_s$ is the net income at time $s$ in state $S_s$, discount factor $\gamma=0.92$, and $T$ is the length of life cycle.  

No saving is included in the model, and consumption in one time-step equals net income.

\subsection{Dynamic programming}
Dynamic programming is a model-dependent method of finding the optimal action policy (Sutton and Barto 2018, Heer and Maussner 2009). 
Dynamic programming is here implemented as a grid method starting from the end state, and working time-step by time-step backwards. The method is also called {\em Value iteration}. To analyze continuous state variables, their values must first be discretized appropriately. Hence, it is also important that the gridß is sufficiently large to provide a good approximation to the problem.

Value iteration requires that a way to connect states at subsequent time points is known (Sutton and Barto 2018). Here, the connecting feature is the probability of getting a certain wage given the previous wage and the current wage. The knowledge of the underlying model is required to obtain this probability distribution. Values in the grid are approximated with cubic splines to improve accuracy.

There are four major state variable directions in which the grid size can vary: {\em time in the state}, {\em wage}, {\em the previous wage} and{\em the accrued pension}. Due to discrete time-step, {\em time in the state} variable is discrete. However, {\em the accrued pension}, {\em wage} and {\em the previous wage} are all essentially continuous variable, and must therefore be discretized in dynamic programming. 
 
Analysis of the convergence suggests that while there may be some minor improvement available by tuning the grid size, the dynamic programming is quite close to the global optimum. Each additional dimension in the state space would multiply the size of the grid by the number of grid points in that direction. 

\subsection{Reinforced learning}
Reinforced learning is a model-free method of solving Markov decision processes. 
Reinforced learning does not use any information on the specifics of the model. It improves its recipe for actions, action policy 
by observing how the state of an agent evolves in response to actions.

Deep reinforced learning algorithms rely on function approximation methods to describe the values of the states and the optimal policy. In principle, any function approximation method can be used, e.g., multivariate regression, decision trees or neural networks, however, in practice neural networks seem to have the best performance (Sutton and Barto 2018).

In a large state space, only a relatively small portion of the entire state space is encountered in training, which requires that the function approximation method is capable of generalizing the action policy from the observed states to those not observed (Sutton and Barto 2018). One of the differences between dynamic programming and reinforced learning algorithms is that while dynamic programming can discriminate well between states, it is not good at generalizing. On the other hand, reinforced learning can generalize, but it is not always clear that the algorithm can discriminate between states requiring different actions.

A good capability to generalize will speed the learning significantly. For extremely large state spaces, this is the only way of solving them in practice. Hence, generalization is a required part of the toolbox for solving realistic life cycle models.

The two main kinds of deep reinforced learning algorithms are value-based algorithms, such as Q-learning (Mnih\etal 2015), and policy gradient methods used here (Sutton and Barto 2018). Both algorithm families assume that Markov decision processes are considered. The main difference between the methods is in whether the aim is to estimate the value of {\em (state,action)} pairs or more directly estimate the action selection policy (Sutton and Barto 2018). 
Action policy refers to the rule connecting a state to an action.
Actor-Critic algorithms are a special case of policy gradient methods, in which the aim is to estimate the optimal action policy. 

Here we use an Actor-Critic algorithm known as ACKTR (Wu\etal 2017), which uses a Kronecker-factored approximation to natural policy gradient allowing the covariance matrix of the gradient to be inverted efficiently (Wu\etal 2017). In our tests, ACKTR converged more rapidly than other Actor-Critic algorithms. Rapid convergence is a desirable feature for life cycle models with a huge state space, even though we are even more interested in obtaining behavior that is sufficiently optimal to be useful for estimating employment rate. 
The results of Wang\etal (2019) suggest that actor-critic algorithms converge towards the globally optimal policy.

To improve convergence of the results, we have applied the usual repertoire of tricks: one-hot encoding of employment states, normalization of wages, ages, times in the state and pensions, scaling of rewards. The policy and value networks use separate artificial neural nets of with two hidden layers. The policy network is of size (32,32,32) and value network is of size  (128,128,128) without the output layers. No regularization was applied to the networks in training. Leaky ReLU was used as the activation function for neural networks.

\subsection{Implementation and performance}
The life cycle model is implemented in the OpenAI Gym framework (Brockman\etal 2016). The used reinforced learning library Stable Baselines 2 (Hill\etal 2018) is compatible with Gym-environments. 
This version of the life cycle model is not aimed at realistically reproducing employment and unemployment rates. It is rather a toy model aimed at 
comparing dynamic programming and deep reinforced learning methods. The entire life cycle model is freely available as {\em unemployment-rev-v0} environment (Tanskanen 2019b). 

While the implementations presented here are not optimized for performance, it may be informative to consider how long the computations take. 
In dynamic programming, a relatively small grid is used. It has 3 employment states, 20 pension accruals, 20 previous wage levels, 25 current wage levels, 5 time-in-state points, and 43 ages, that is, $3\times 20\times 20\times 25\times 5\times 43$ grid points. Grid is further interpolated with cubic splines. It takes about 4.5 hours to fit the model on an Apple MacBook Pro (2019). Training ACKTR for 10 million time-steps requires about 1.5 hours. 

For larger grid sizes, it takes combinatorially longer to solve the model with DP. For example, the model used in Tanskanen (2020a) contains approximately $10^{21}$ states when appropriately discretized, while the model considered here contains $6.45\cdot 10^6$ states. The number of states in the more complex model of Tanskanen (2020a) renders the use of dynamic programming, as considered here, unfeasible even with considerable computational resources. ACKTR was used to analyze the model in Tanskanen (2020a), and it took roughly 8 hours and 20 million time-steps to fit the model.

Once the model is fitted, behavior of a population of agents must be simulated. With a population of 100,000 agents, it takes 32 minutes with ACKTR and 85 minutes with DP to produce the aggregate results. The implementation of DP could be significantly improved, however, scaling properties would not be improved unless the method was revised. The time it takes to train and simulate ACKTR depends heavily on the size of policy and value networks.

All aggregate results in reinforced learning are averaged over ten simulation runs with 50,000 agents each. Reinforced learning model was fitted 10 times and only then was the life cycle of the 50,000 agents predicted. In the dynamic programming model, fitting was done only once, and simulation was run on a population of 100,000 agents.

\section{Computational results}
\begin{figure}
\includegraphics[width=7.5cm]{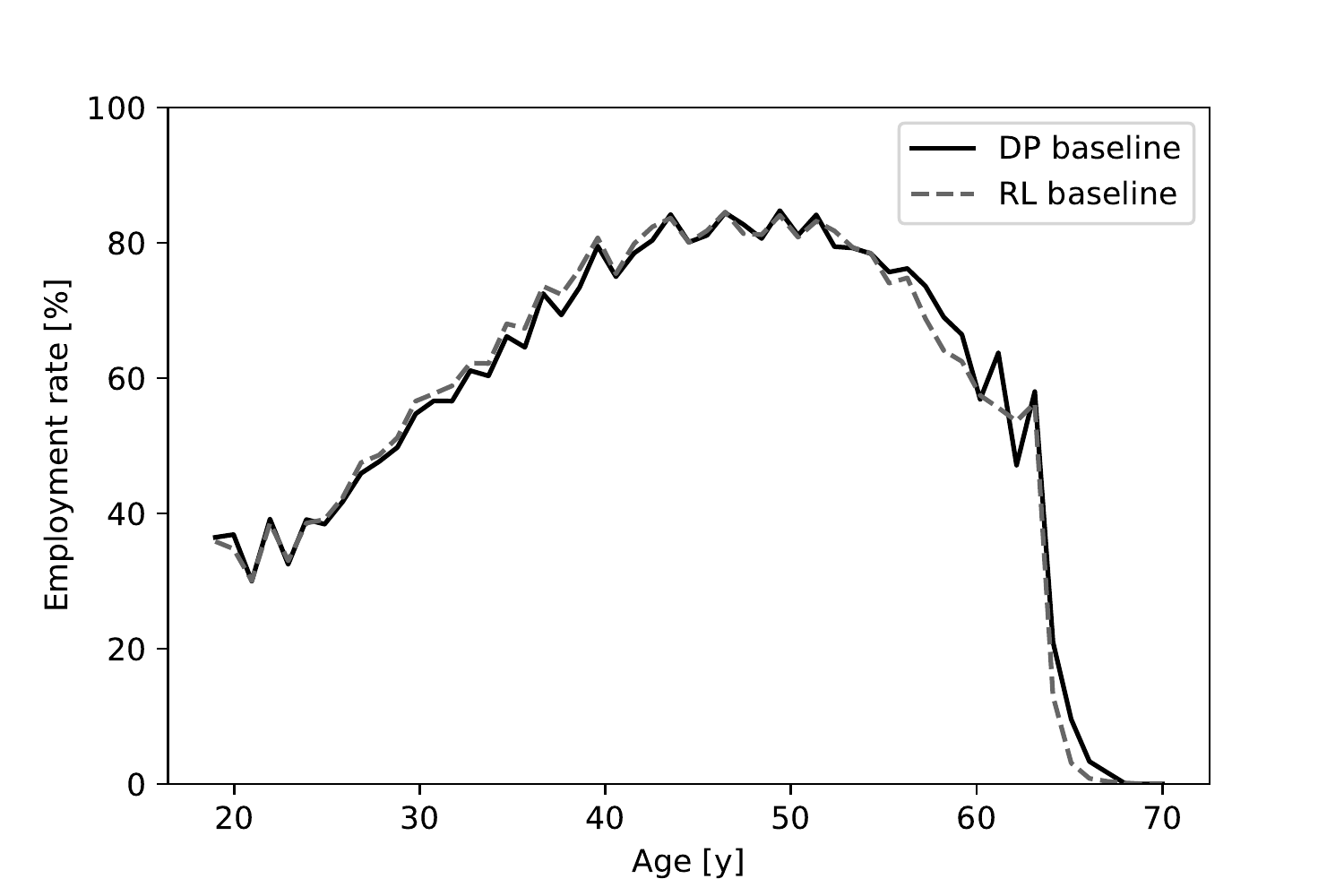}
\includegraphics[width=7.5cm]{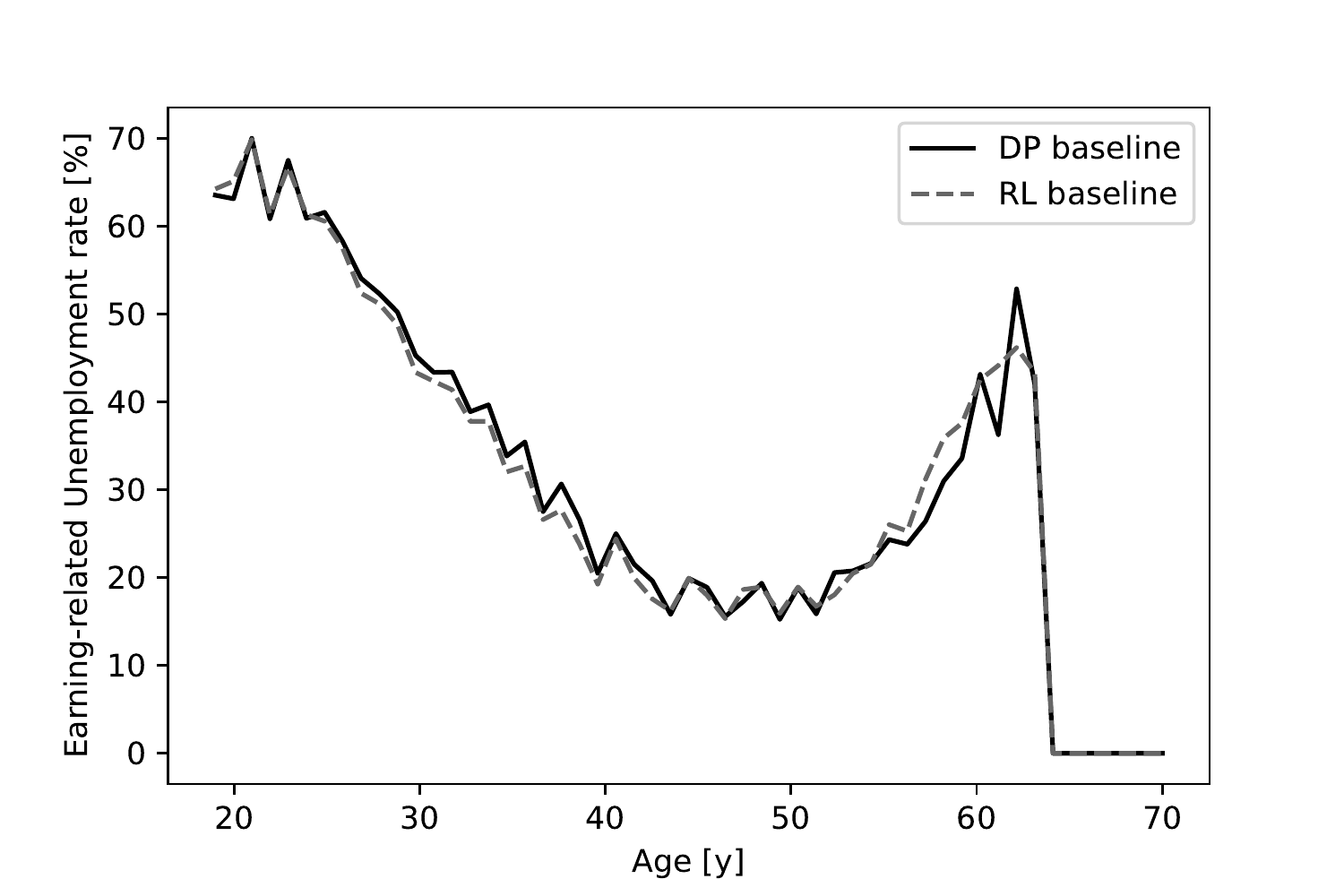}
\caption{Employment rate (left panel) and unemployment rate (right panel) in the baseline life cycle model. Panels compare the rate solved with dynamic programming (DP; black solid line) and in reinforced learning algorithm (RL; gray dashed line). }
\label{fig:fig2}
\end{figure}

\subsection{Discounted utility}
Table 1 shows the best results obtained with dynamic programming and compares them against those found using reinforced learning algorithm ACKTR and random selection of actions. In a life cycle model, the object of maximization is the {\em expected discounted utility} in equation (\ref{eq:1}) at each time-step. Expected discounted utility provides a natural measure to compare the results from various algorithms\footnote{The most straight forward way would be to compare the undiscounted sum of period utilities per agent obtained by each algorithm. However, it is a measure that is relatively hard to interpret and it is not an aim of optimization in a life cycle model. Therefore, it would not give much information on whether dynamic programming produced better results than reinforced learning, or visa versa.}. Table 1 shows the expected discounted utility at the start of simulation (column {\em initial} discounted utility). Dynamic programming yields 0.001 higher initial expected discounted utility than reinforced learning.  

\begin{table}
  \begin{center}
    \begin{tabular}{|c|rrr|}
    \hline
      & DP & RL  & random\\
    \hline
    Initial discounted utility & 11.775 & 11.774 & 11.687 \\
    Average discounted utility & 11.122 & 11.121 & 10.853 \\
    Compensating consumption (\%) & - & 0.081 & 7.39 \\
    Equivalent net income (e/y) & 13,209.22 & 13,151.61 & 11,434.57\\
    Employment & 2,083,615 & 2,075,449 & 1,664,799\\
    \hline
    \end{tabular}
  \end{center}
  \caption{Comparison of the results from dynamic programming (DP), reinforced learning (RL), and random action selection (random).}
  \label{table:1}
\end{table} 

In addition to initial discounted utility, Table 1 shows the {\em time-averaged discounted utility}. However, it is a path-dependent measure and therefore does not directly show which method is better. 
The time-averaged discounted utilities obtained with dynamic programming and reinforced learning are close to each other in the baseline (Average discounted utility in Table 1). The results are compared to those obtained by an uninformed agent (random in Table 1) that takes random actions at each time point.  Dynamic programming produces 0.001 higher time-averaged discounted utility than reinforced learning, and 0.269 higher than uninformed agent.

The resulting employment rates are quite close to each other (Fig. 1). Dynamic programming forecasts a slightly higher employment than reinforced learning (Table 1) and a slightly lower unemployment (Fig. 1). The main difference between the employment rates is the presence of spikes in the employment rate forecasted by DP (Fig. 1). The spikes are a result of a large number of agents acting in a similar way: one year of working enables one year of earnings-related unemployment benefit but further earnings-related periods require that at least 1 year of working precedes the unemployment. According to DP, this is the optimal behavior near the lowest retirement age, however, RL seems to miss the spikes.

\subsection{Equivalent net income}
{\em Compensating consumption} is here defined as the average additional consumption that would make the initial discounted utility in RL 
equal to that obtained with DP (as in Cooley and Soares, 1999)\footnote{More precisely, compensating consumption is defined here as a solution of $x$ in equation 
\begin{equation*}
\frac{1}{N}\sum_{i=1}^N\sum_{t=1}^T\beta^{t-1}u(c_{t,i}^{DP},l_{t,i}^{DP})=\frac{1}{N}\sum_{i=1}^N\sum_{t=1}^T\beta^{t-1}u((1+x)c_{t,i}^{RL},l_{t,i}^{RL}).
\end{equation*}
}. 
The consumption in RL would need to be increased by 0.081 percent to equal the consumption in DP. In the random policy, the increase would need to be 7.39 percent of consumption. 

{\em Equivalent net income} describes the time-averaged net income adjusted by the value of lost free time. This is an undiscounted measure and as such does not directly correspond to the target of optimization.
Dynamic programming gives a somewhat higher average equivalent net income than reinforced learning (Table 1). 
Again, the uninformed, randomly acting agent gets a significantly lower average equivalent net income.

\subsection{Optimal action policy}

\begin{figure}
\includegraphics[width=7.5cm]{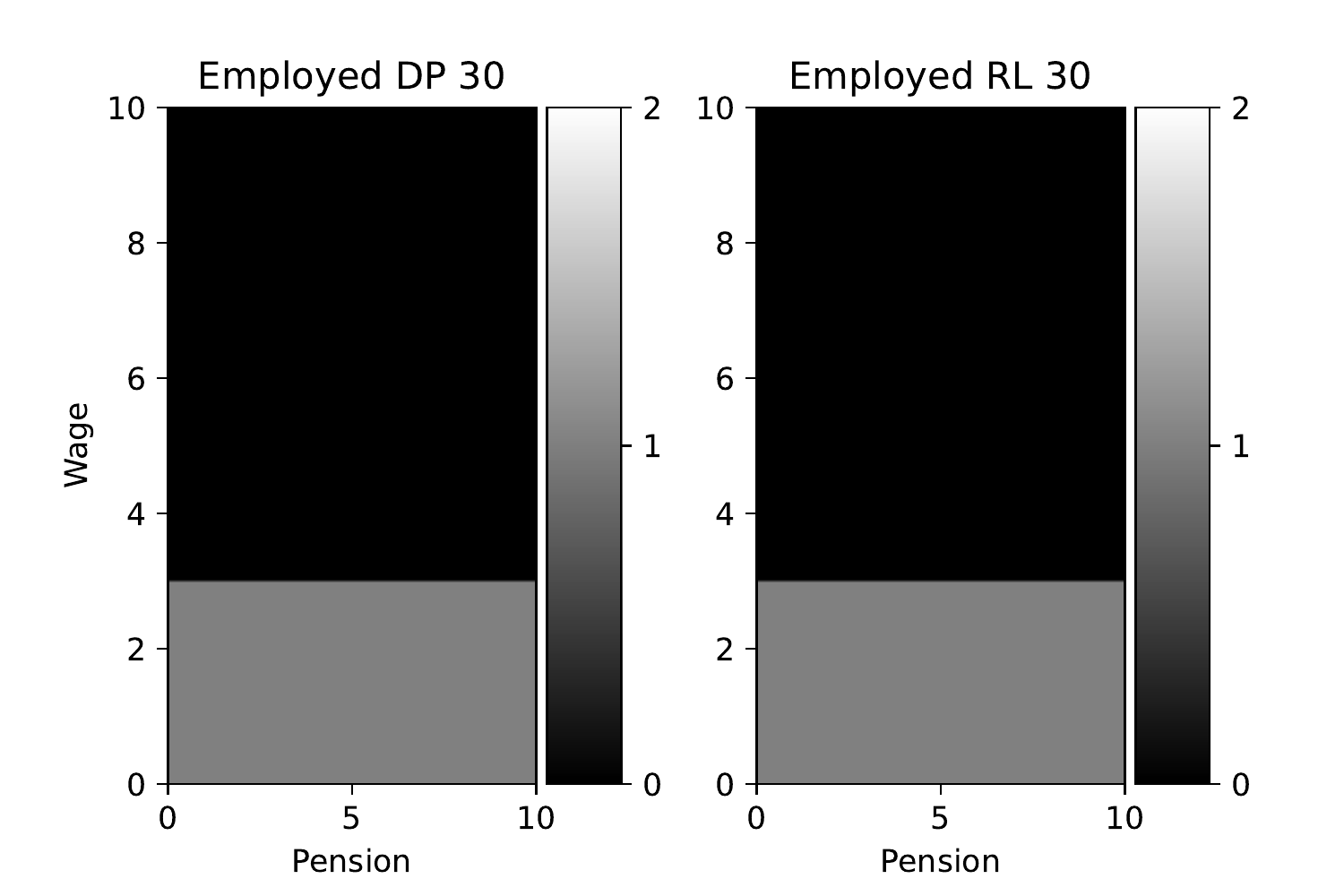}
\includegraphics[width=7.5cm]{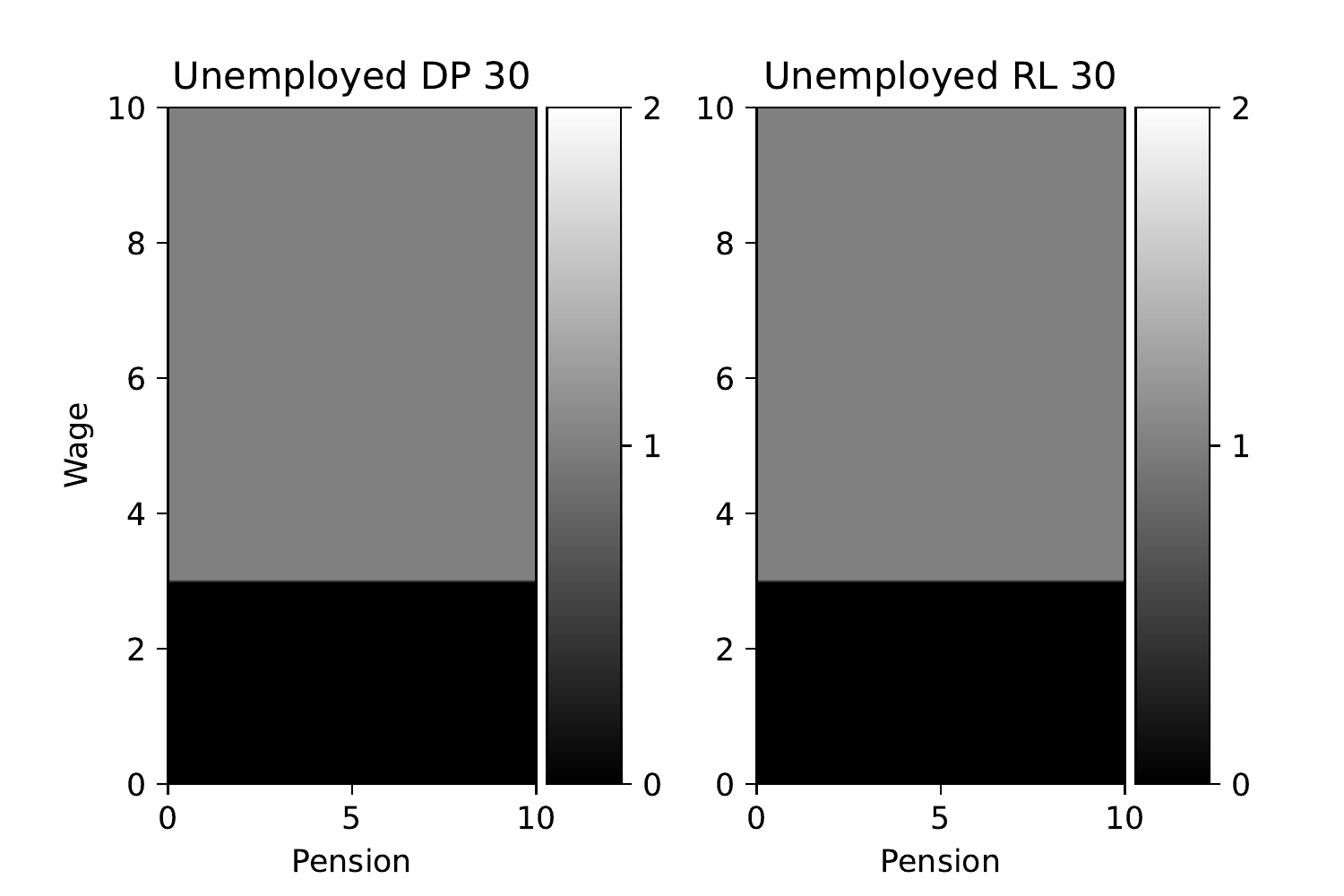}
\includegraphics[width=7.5cm]{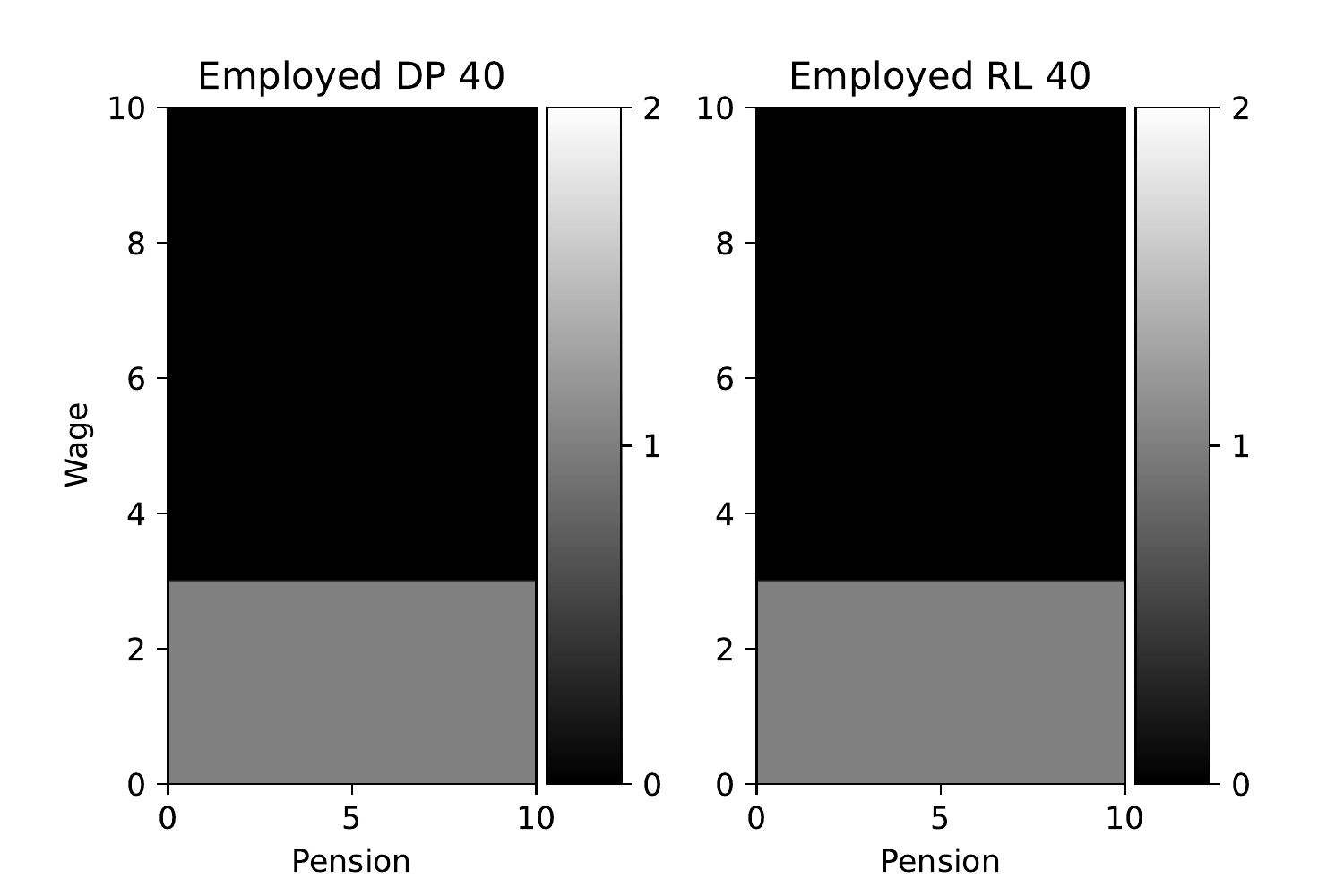}
\includegraphics[width=7.5cm]{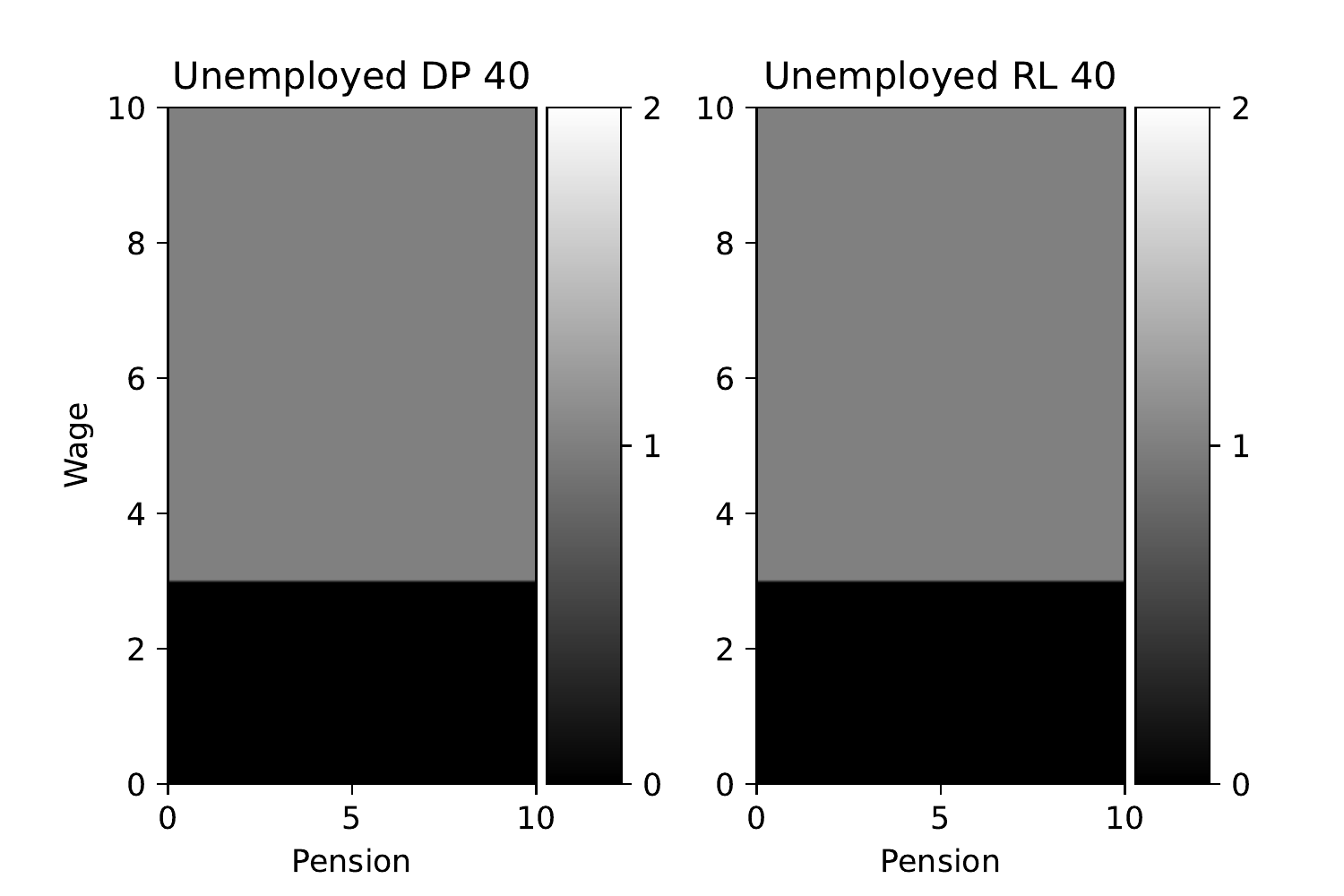}
\includegraphics[width=7.5cm]{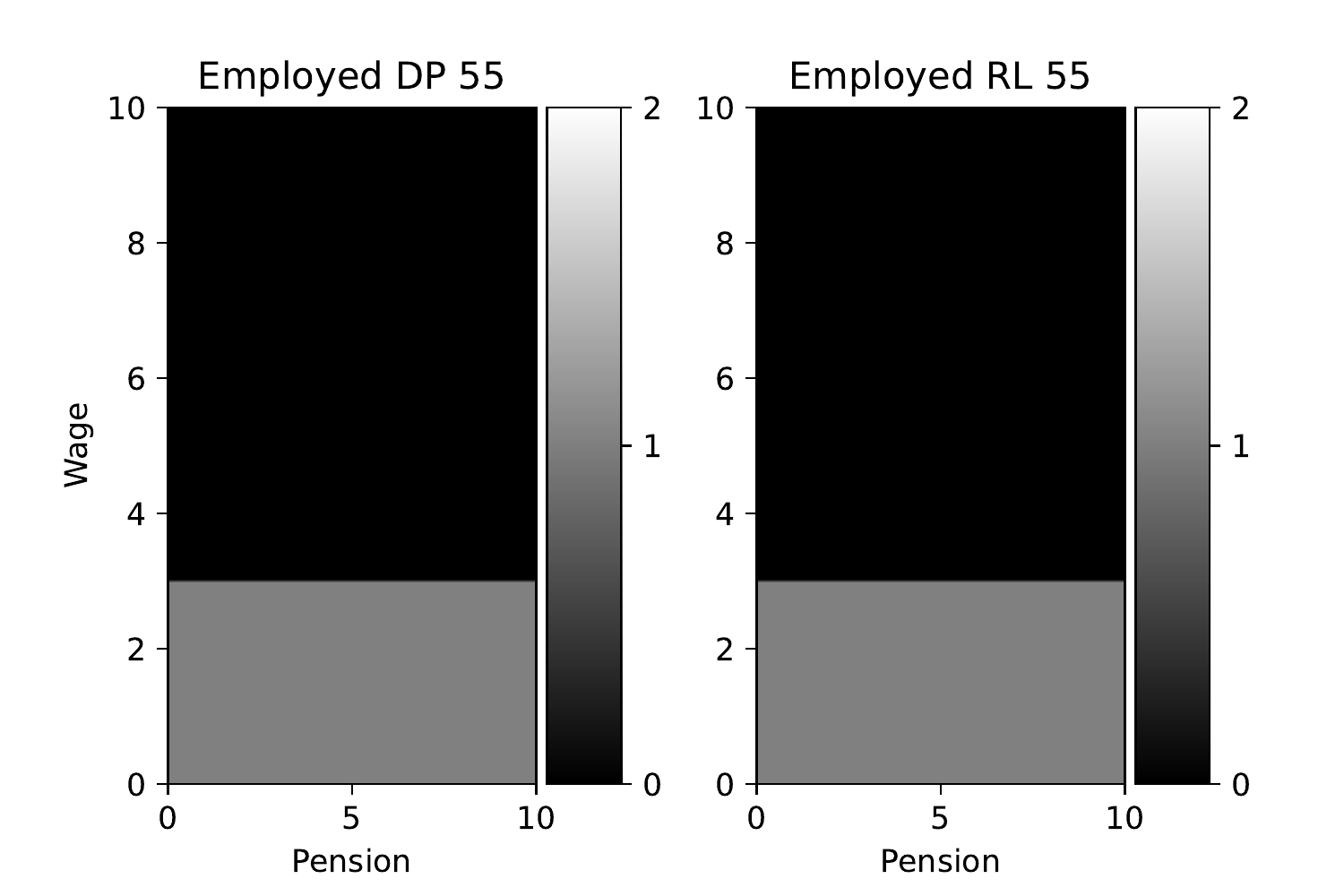}
\includegraphics[width=7.5cm]{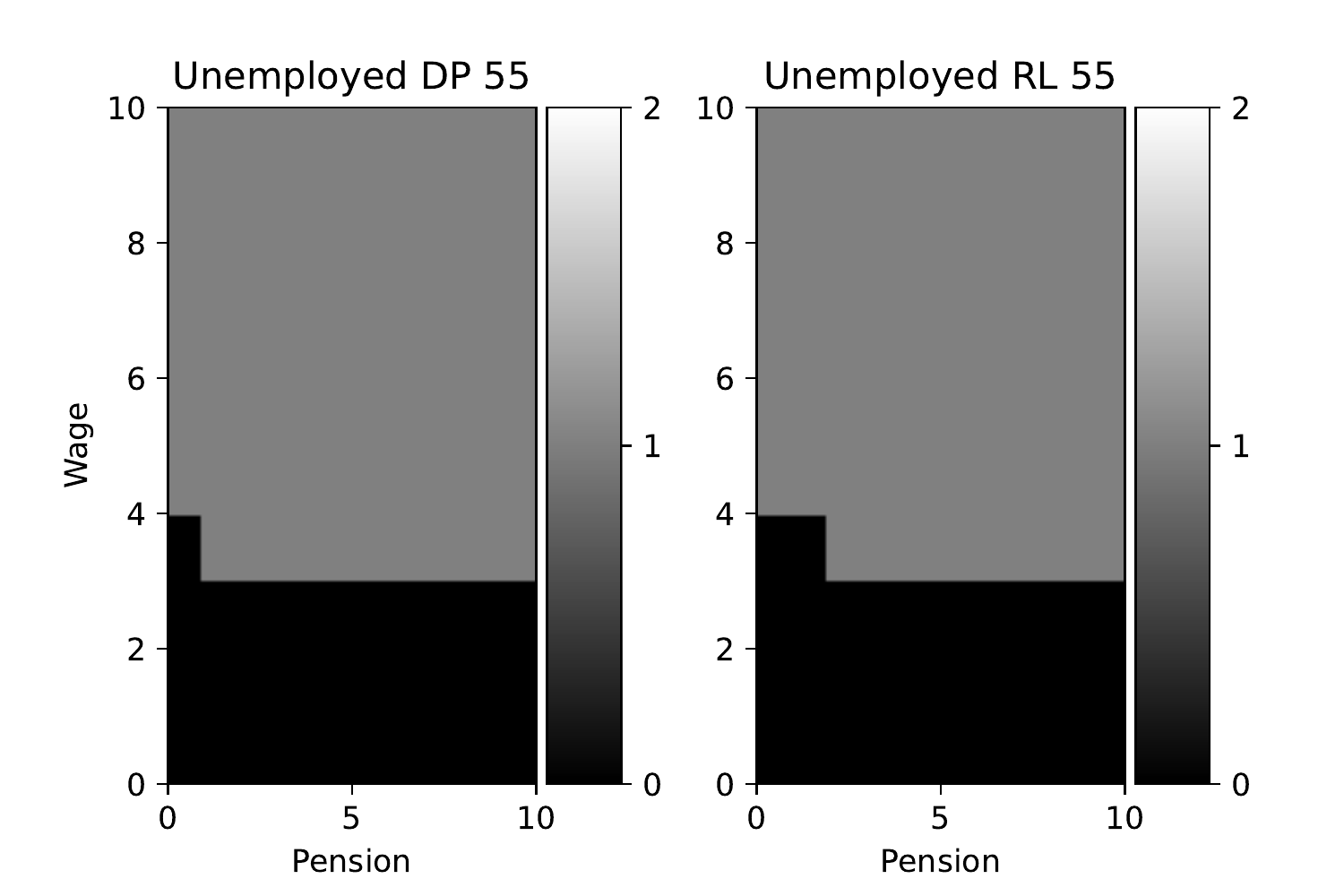}
\caption{Optimal actions at ages of 30, 40 and 55 solved with reinforced learning algorithm (RL) and dynamic programming (DP). The two left columns describe agents in state employed, while the two right columns describe agent in state unemployed. Horizontal axis is in each case the accrued pension, while the vertical axis is the salary. Action 0 (black) stays in the current employment state, Action 1 (gray) switches between employed and unemployed, Action 2 (white) describes retiring. The axes' labels refer to grid points in dynamical programming.}
\label{fig:fig4}
\end{figure}

The aggregate employment rates in Fig. 1 and the statistics in Table 1 do not tell how optimal the found action policies in various states differ from each other in RL and in DP. 
Figures 2 and 3 show the optimal action policies in dynamic programming and reinforced learning. Each color represents an action: Action 0 (black squares) depicts staying in the current employment state, Action 1 (gray squares) depicts switching between employed and unemployed (when employed) or between unemployed and employed (when unemployed), Action 2 (white squares) depicts retiring. The axis labels refer to grid points in dynamic programming. The grid starts from 0 euros/month in pensions and from 83 euros/month in wages. Each step in wages is 701 euros/month and in pensions 417 euros/month.

Reinforced learning reproduces the optimal actions of dynamic programming in the employed state almost identically until the retirement age (Fig. 2 and Fig. 3). At age 64, there is a slightly different boundary between retiring and staying employed (Fig. 3).

When unemployed, the optimal action policies differ slightly more than in the employed state. The differences begin at age 55, where the boundary between staying unemployed and switching to employment has a slightly different shape (Fig. 2). Similar issues are present at ages 60 and 62 (Fig. 3). 
At the lowest retirement age, the boundary between switching to employment and retiring is somewhat different in DP and in RL.

Fig. 4 shows two examples of the sampled {\em (wage,pension)} pairs in reinforced learning. The points are relatively tightly concentrated, which suggests that reinforced learning learns from a quite limited dataset. The action policy for states that were not visited during the training is generalized from the action policy learned from the visited states. Still, reinforced learning solves the baseline model quite well.
\begin{figure}
\includegraphics[width=7.5cm]{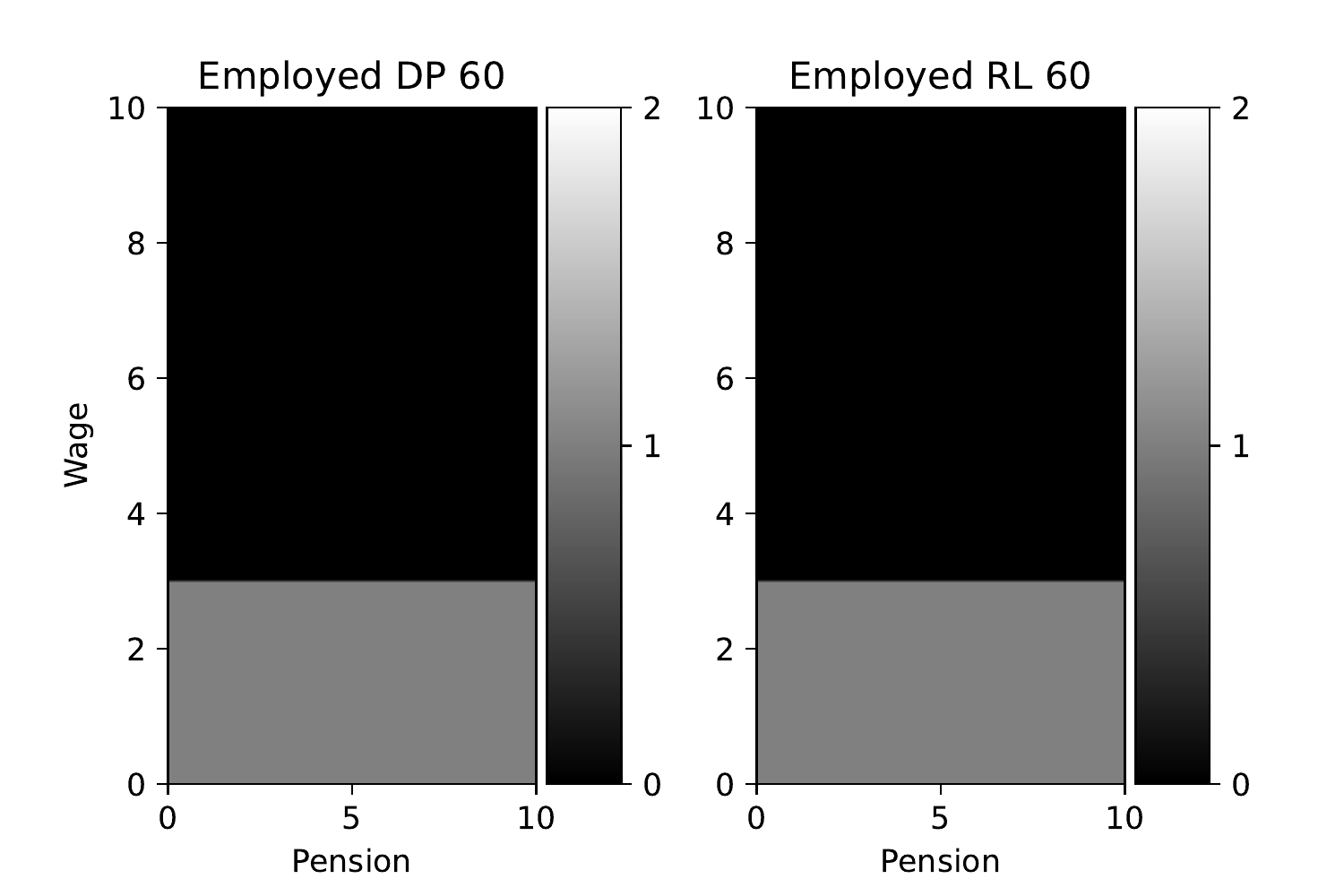}
\includegraphics[width=7.5cm]{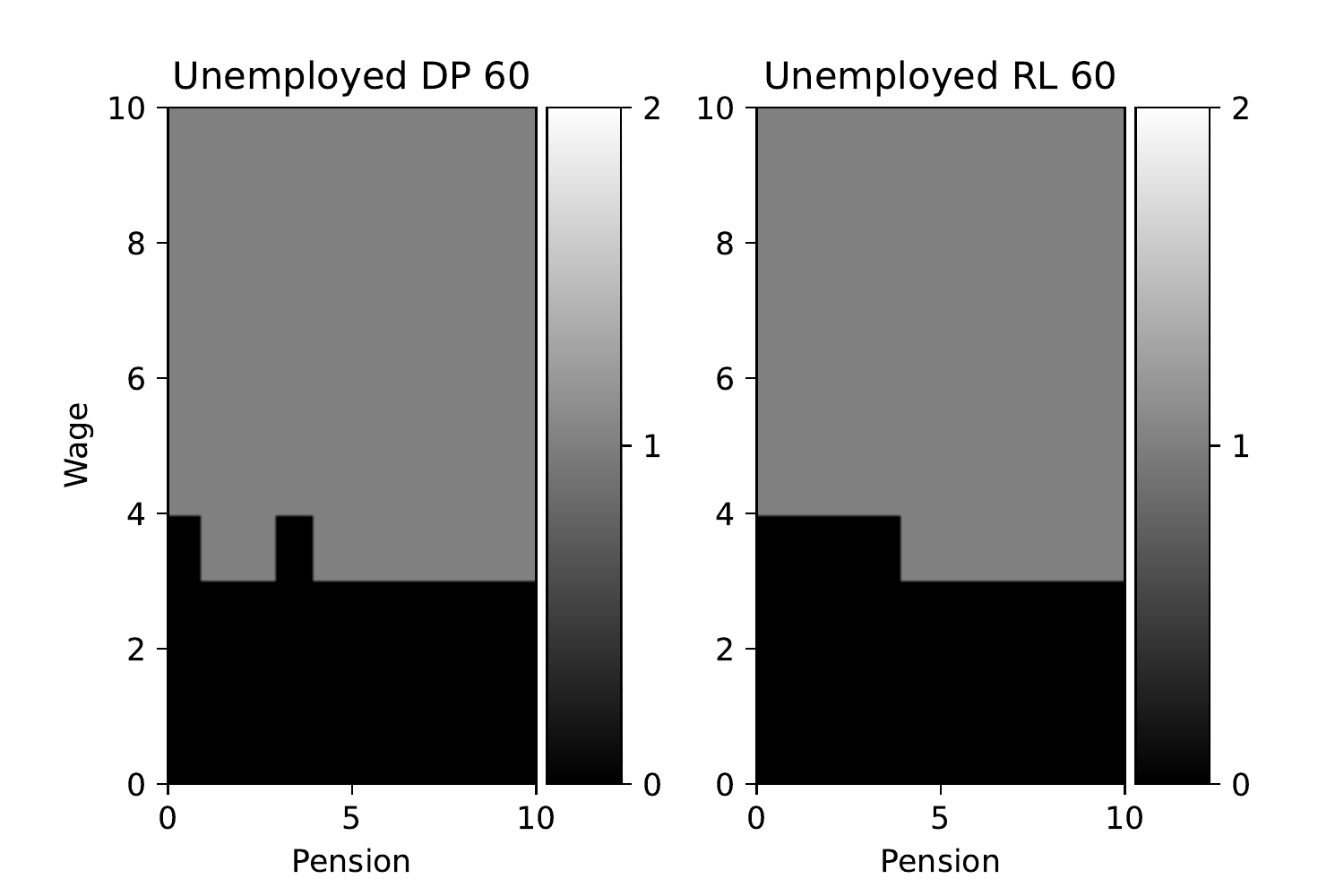}
\includegraphics[width=7.5cm]{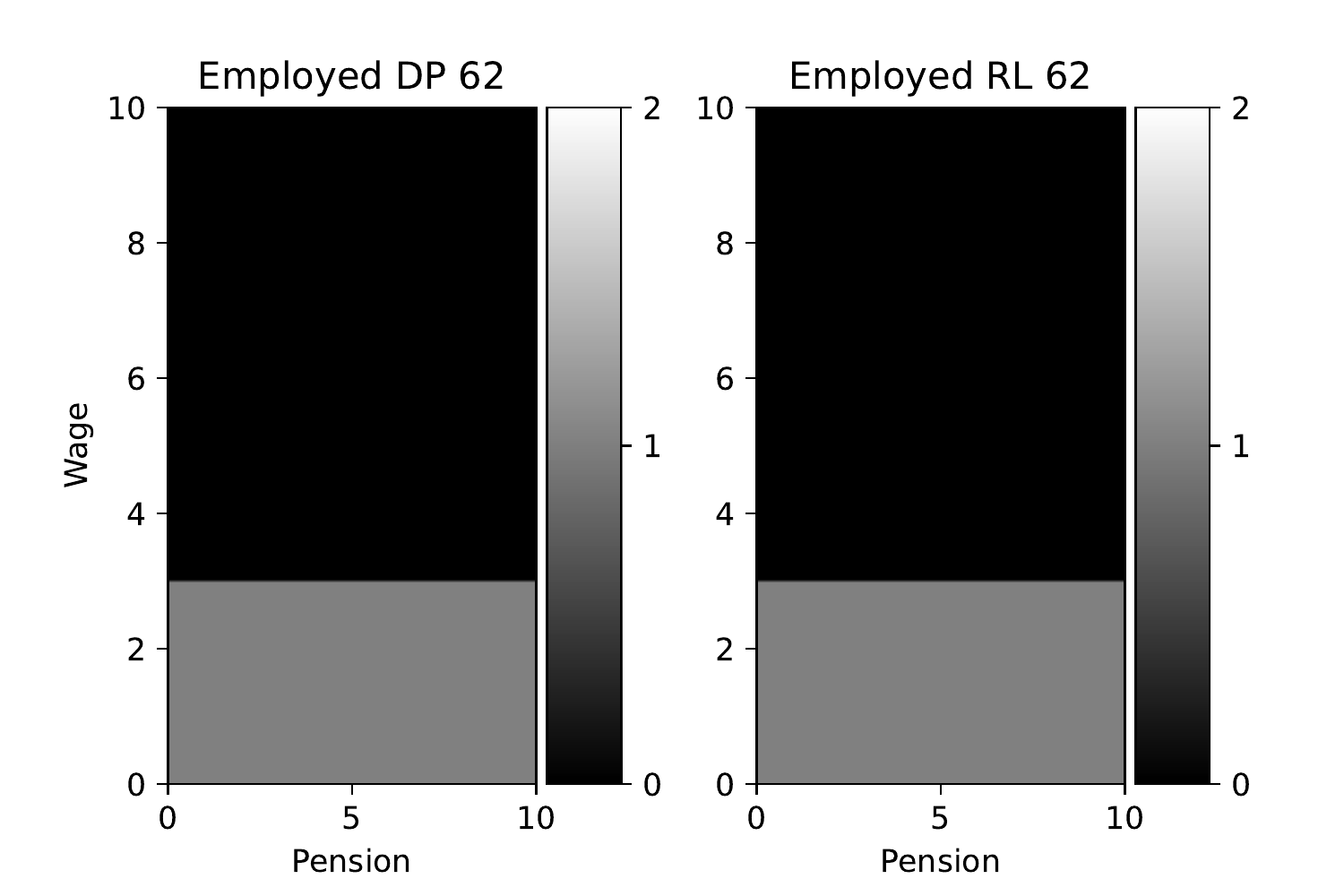}
\includegraphics[width=7.5cm]{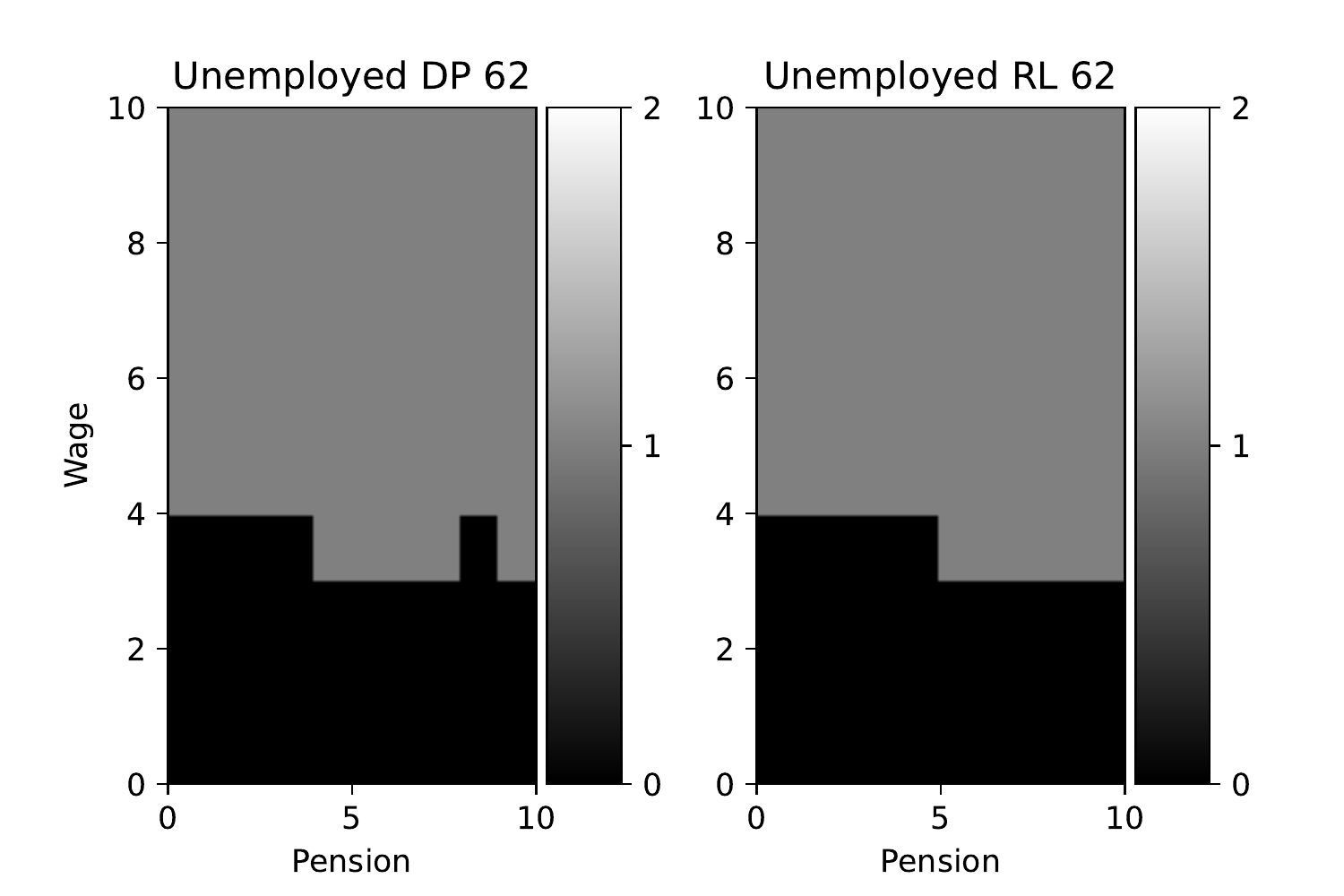}
\includegraphics[width=7.5cm]{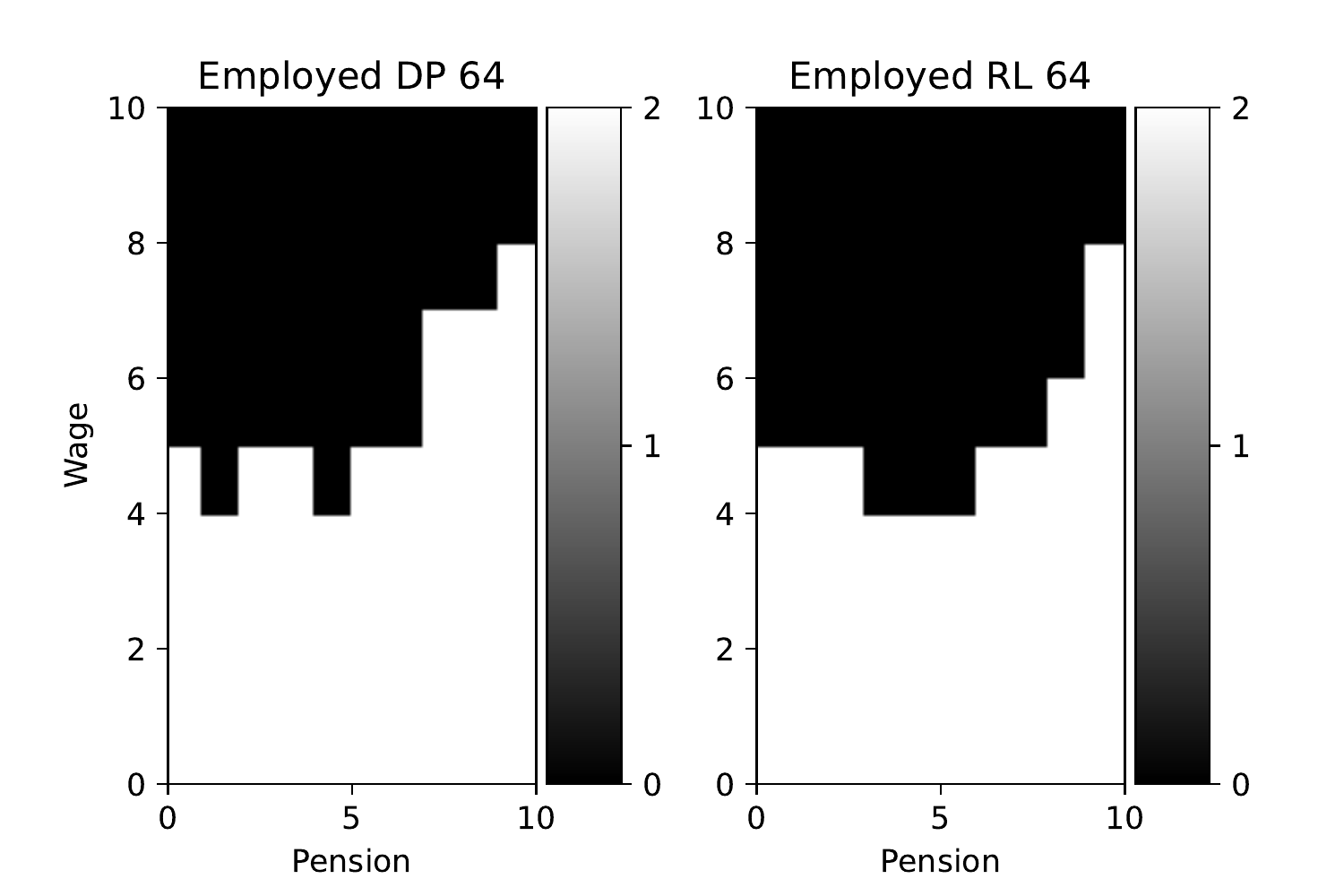}
\includegraphics[width=7.5cm]{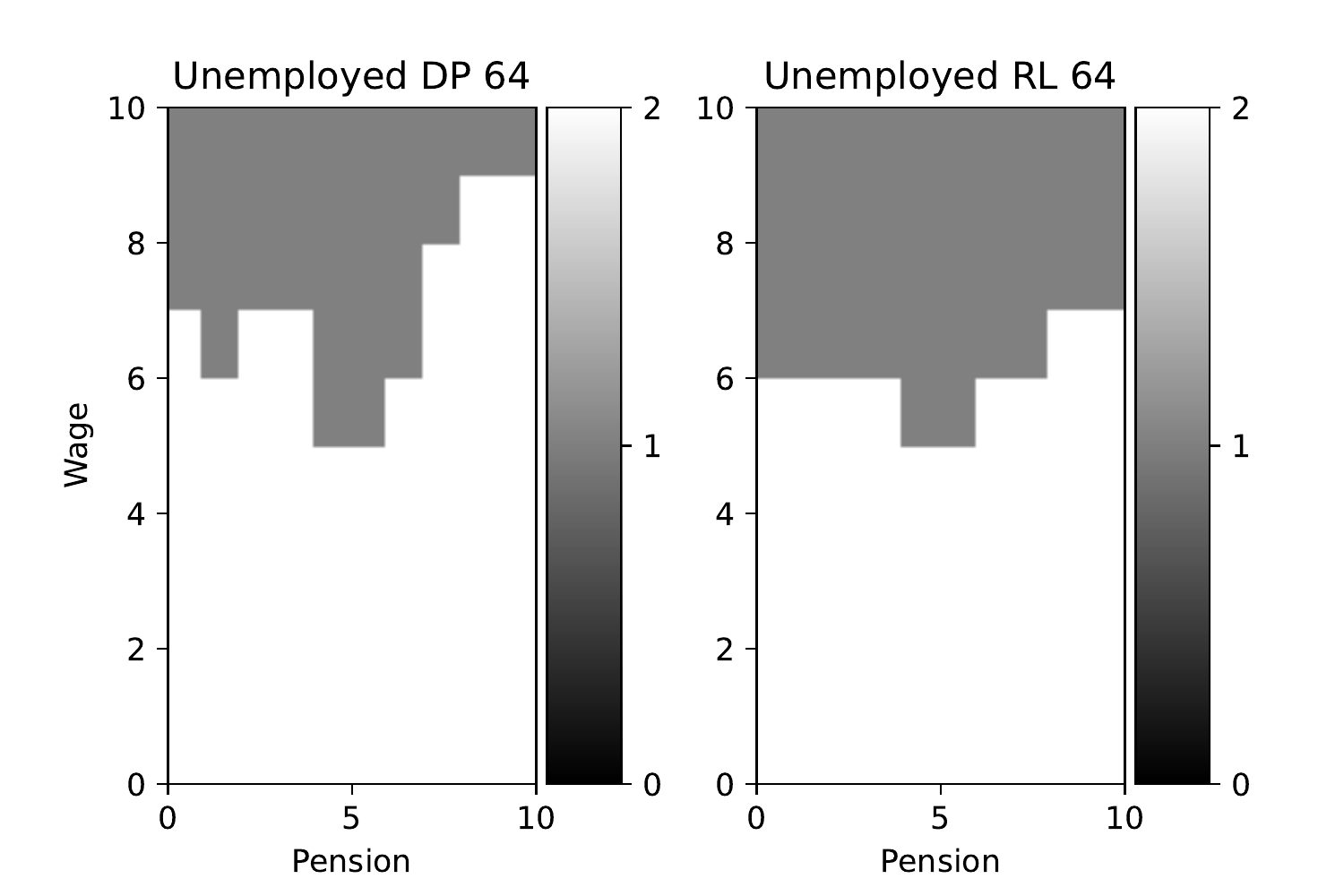}
\caption{Optimal actions at ages of 60, 62 and 64 solved with reinforced learning algorithm (RL) and dynamic programming (DP). The two left columns describe agents in state employed, while the two right columns describe agent in state unemployed. Horizontal axis is in each case the accrued pension, while the vertical axis is the salary. Action 0 (black) stays in the current employment state, Action 1 (gray) switches between employed and unemployed, Action 2 (white) describes retiring. The axes' labels refer to grid points in dynamical programming.}
\label{fig:fig5}
\end{figure}

\begin{figure}
\includegraphics[width=7.5cm]{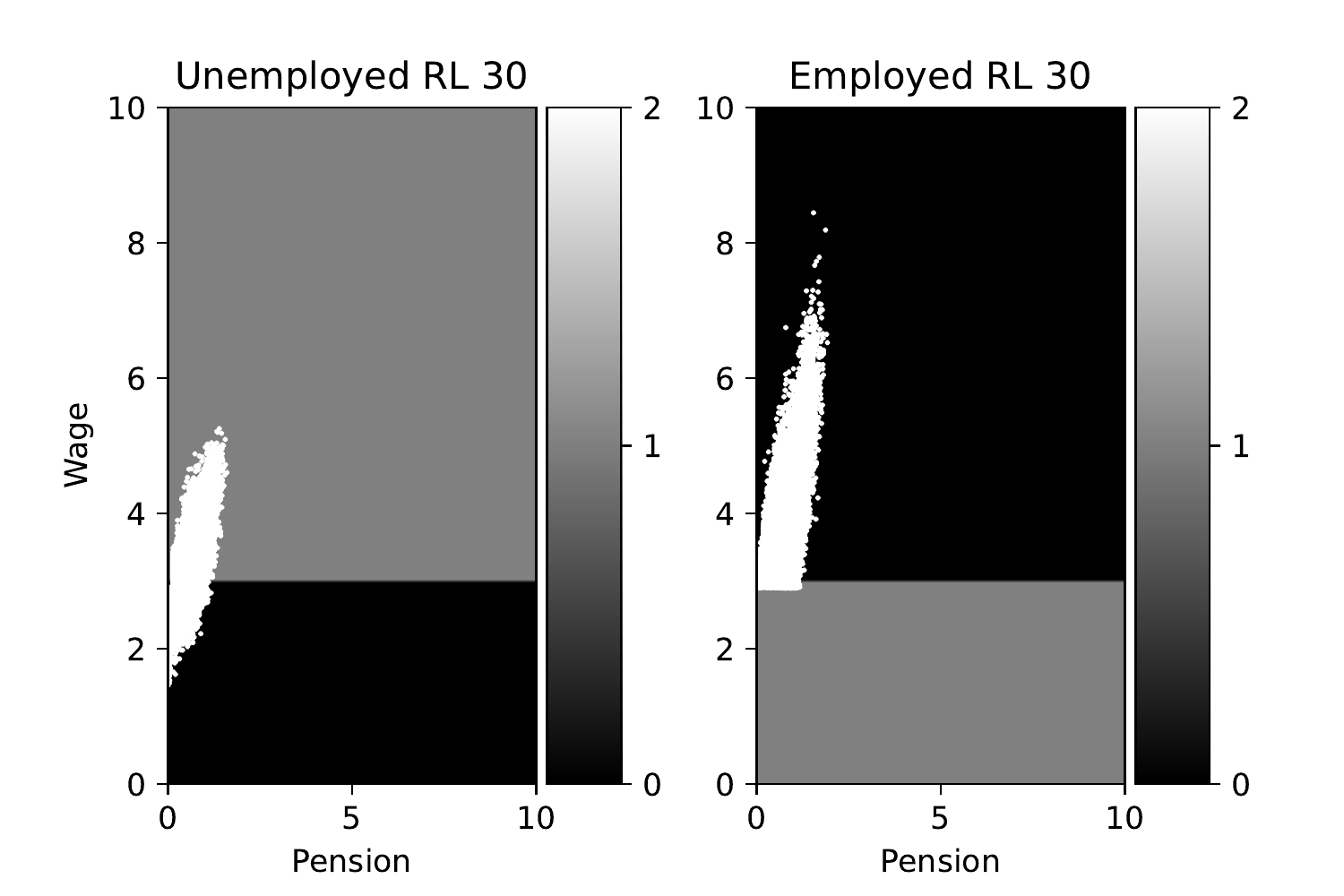}
\includegraphics[width=7.5cm]{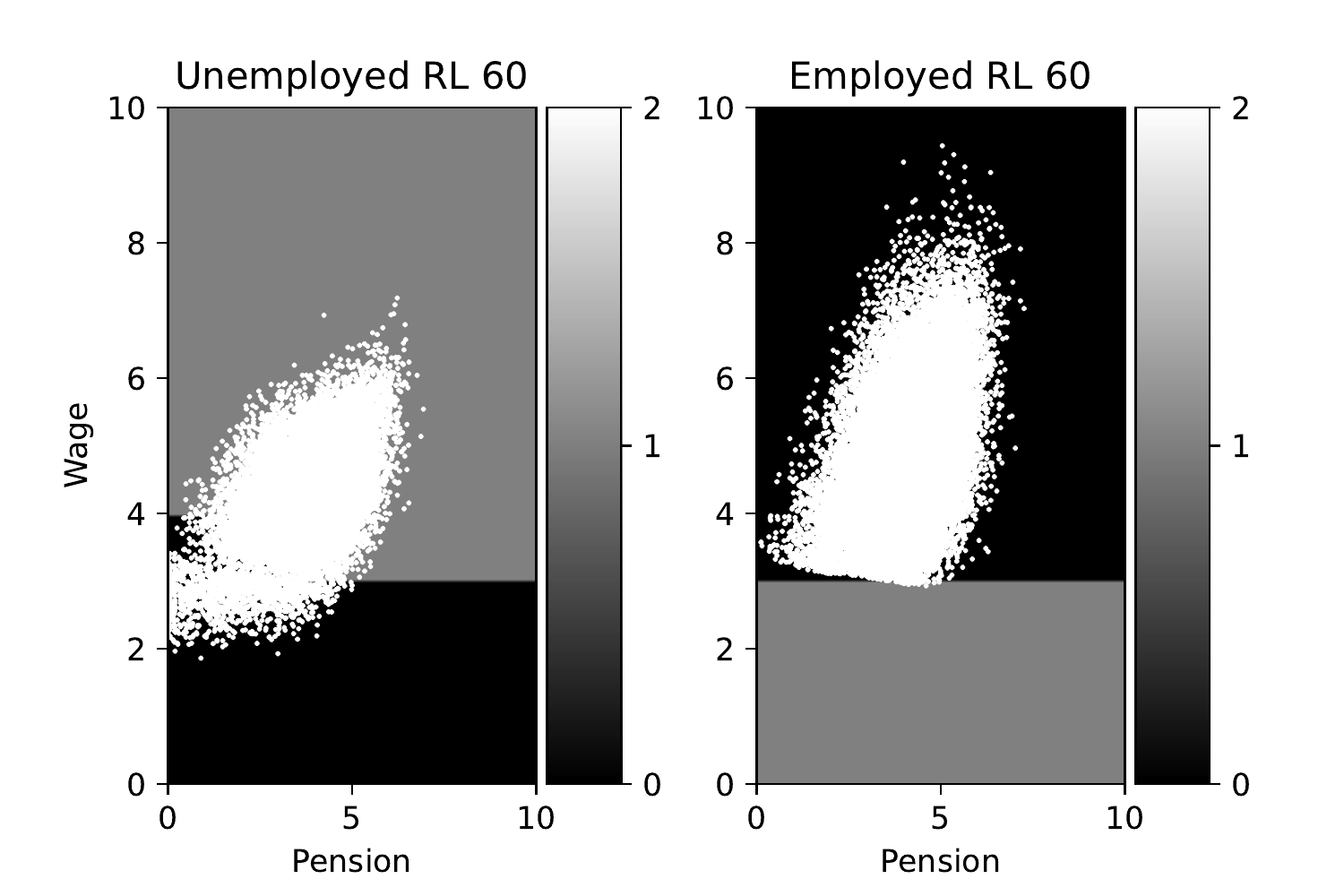}
\caption{An example of 50,000 (wage, accrued pension) pairs (dots in white) observed in simulation. The observed pairs are plotted against the optimal actions in reinforced learning.}
\label{fig:fig6}
\end{figure}

\section{Policy reforms}

To find out how the algorithms predict the impact of a policy reform on employment, we consider two reforms: (1) an increase of the retirement age from 63.5 to 66 years, and (2) a universal basic income at 500 euros per month. The aim is to find out whether the methods agree or disagree on the impacts of the reforms. 

 \begin{table}
  \begin{center}
    \begin{tabular}{|c|r|rrr|}
    \hline
    Reform & Statistic & DP & RL & $\Delta$\\
    \hline
    Baseline & Initial discounted utility & 11.775 & 11.774 & 0.001 \\
    & Time-averaged discounted utility & 11.122 & 11.121 & 0.001 \\
    & Equivalent net income (e/y) & 13,209.22 & 13,151.61 & 57.61\\
    & Employment & 2,083,615 & 2,075,449 & 8,166\\
    \hline
    Retirement age & Initial discounted utility & 11.773 & 11.769 & 0.004 \\
    & Time-averaged discounted utility & 11.106 & 11.104 & 0.002 \\
    & Compensating consumption (\%) & 0.19 & 0.10 & 0.09\\
    & Equivalent net income (e/y) & 13,033.32 & 12,998.62 & 34.70\\
    & Employment & 2,188,574 & 2,169,933 & 18,641\\
    & Impact on employment & 104,959 & 94,484 & 10,475 \\
    \hline
    UBI & Initial discounted utility & 11.392 & 11.391 & 0.001 \\
    & Time-averaged discounted utility & 10.773 & 10.772 & 0.001 \\
    & Compensating consumption (\%) & 36.09 & 36.04 & 0.05\\
    & Equivalent net income (e/y) & 9,863.68 & 9,843.56 &  2.02 \\
    & Employment & 1,600,156 & 1,620,632 & -20,000 \\
    & Impact on employment & -483,459 & -454,817 & -28,642 \\
        \hline
    \end{tabular}
  \end{center}
  \caption{Statistics of the baseline, the retirement age reform and the universal basic income (UBI) reform. The columns show the results obtained with dynamic programming (DP) and reinforced learning (RL), and the difference between the results ($\Delta$). Compensating consumption is here computed against Baseline.}
  \label{table:2}
\end{table} 

\subsection{Retirement age}

Life span has increased significantly in the recent decades, as well as the ability to work until older ages.
Pension reforms have in many countries, including Finland, increased the retirement age, which has resulted increases in the expected retirement ages in the direction of the public pension reforms, the length of work careers and in the labor force participation among older workers (Pilipiec~\etal 2020). 

Määttänen (2014) demonstrated that an increase in the minimum retirement age increases employment, and that its magnitude can be analyzed in a life cycle model. In a similar fashion, here we consider increasing the minimum retirement age from 63.5 to 66. The reform increases employment by 104,959 person-years in DP and by 94,484 person-years in RL (Table 2). The standard deviation of employment in RL  is 17,785 person-years. Both methods predict that increasing the retirement age prolongs careers.

The retirement age reform would increase employment rate from 50 years (Fig. 5) until the minimum retirement age. DP predicts a slightly higher employment rate than RL at ages above 58 (Figures 5 and 6). At ages above 60, RL predicts a smooth employment rate, while DP predicts more rapid switches between employment and unemployment (Figures 5 and 6). Other than the spikes in employment, the agreement between the results of DP and RL is good.

According to the results, increasing the minimum retirement age would reduce welfare. In DP, welfare is reduced by 0.016 and in RL by 0.017, as measured by time-averaged discounted utility. However, for simplicity we did not change the life expectancy coefficient or pension premiums in the retirement age reform. If the life expectancy coefficient was changed to take the change in the lowest retirement age into account, the paid pensions would be increased further in the reform. If the reform was made cost-neutral, pension premiums should be reduced, which would further increase welfare. Taking the issues into account, the influence of retirement age reform on welfare would be a more positive than shown in Table 2.

 \begin{figure}
\includegraphics[width=7.5cm]{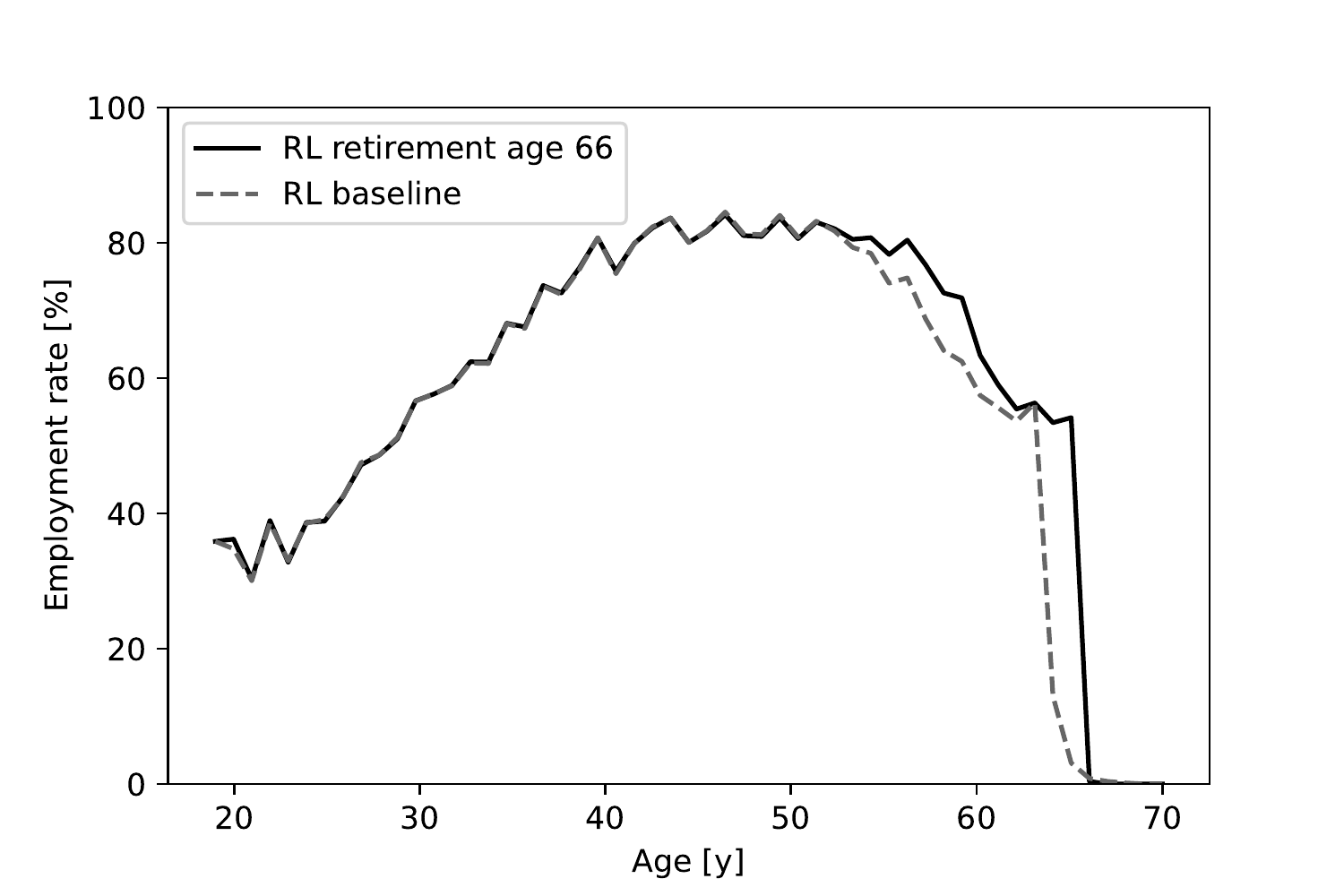}
\includegraphics[width=7.5cm]{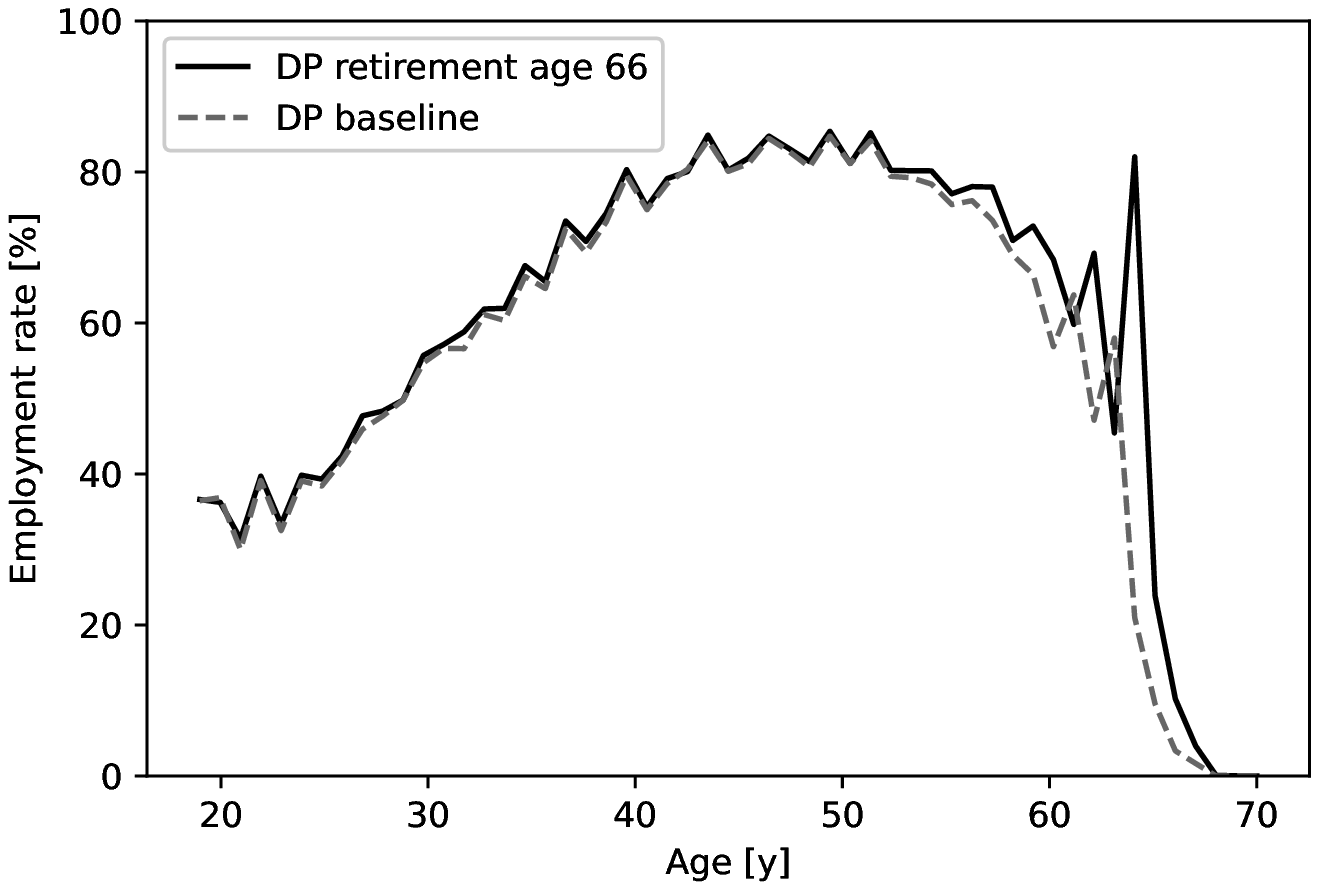}
\caption{Impact of increasing retirement age to 66 years on the employment rates. The reformed model (solid black line) compared to the baseline case (gray dashed line) (A) in reinforced learning, and (B) in dynamic programming. }
\label{fig:fig6}
\end{figure}

 \subsection{Universal basic income}

Universal basic income (UBI) is one of the reforms proposed to address issues in the current social security schemes (Kangas~\etal 2019, De Wispelaere~\etal 2019). Consequences of the adaption of universal basic income depend heavily on the particulars of the reform, such as which benefits UBI replaces and how large UBI is. UBI is often coupled with a flat rate tax, which may be quite high to keep the reform self-financing.

Here UBI at 500 e/month level is considered. The reform is partial in the sense that earning-related social benefits are not abolished, and UBI only replaces the minimum benefits. UBI is not means-tested, and other forms of income do not reduce it. To make the reform cost-neutral, the flat rate tax should be quite high, which reduces the employment rate. A flat rate tax is imposed at 40 percent level. Social security contributions are collected like they are collected currently. 

 \begin{figure}
\includegraphics[width=7.5cm]{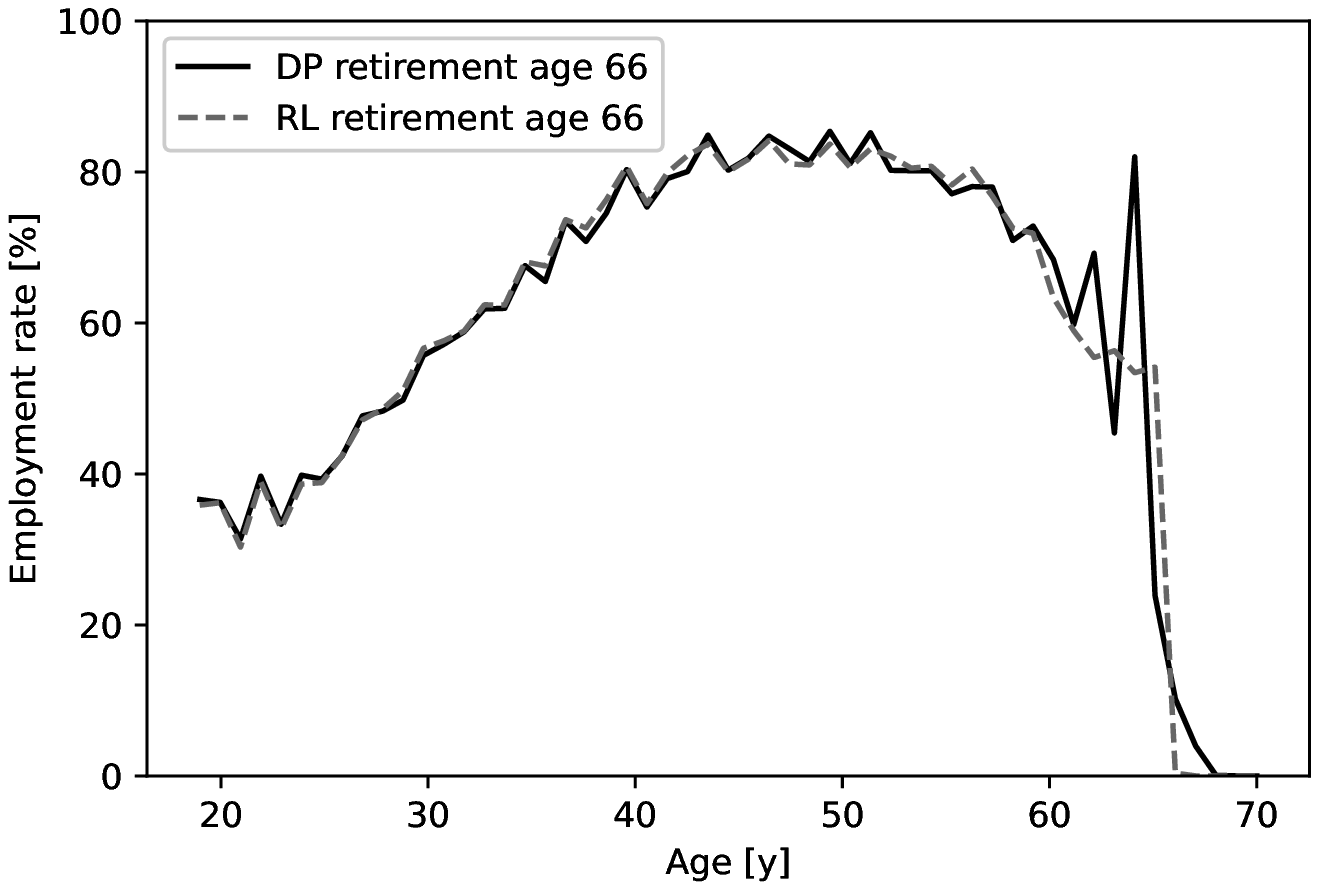}
\includegraphics[width=7.5cm]{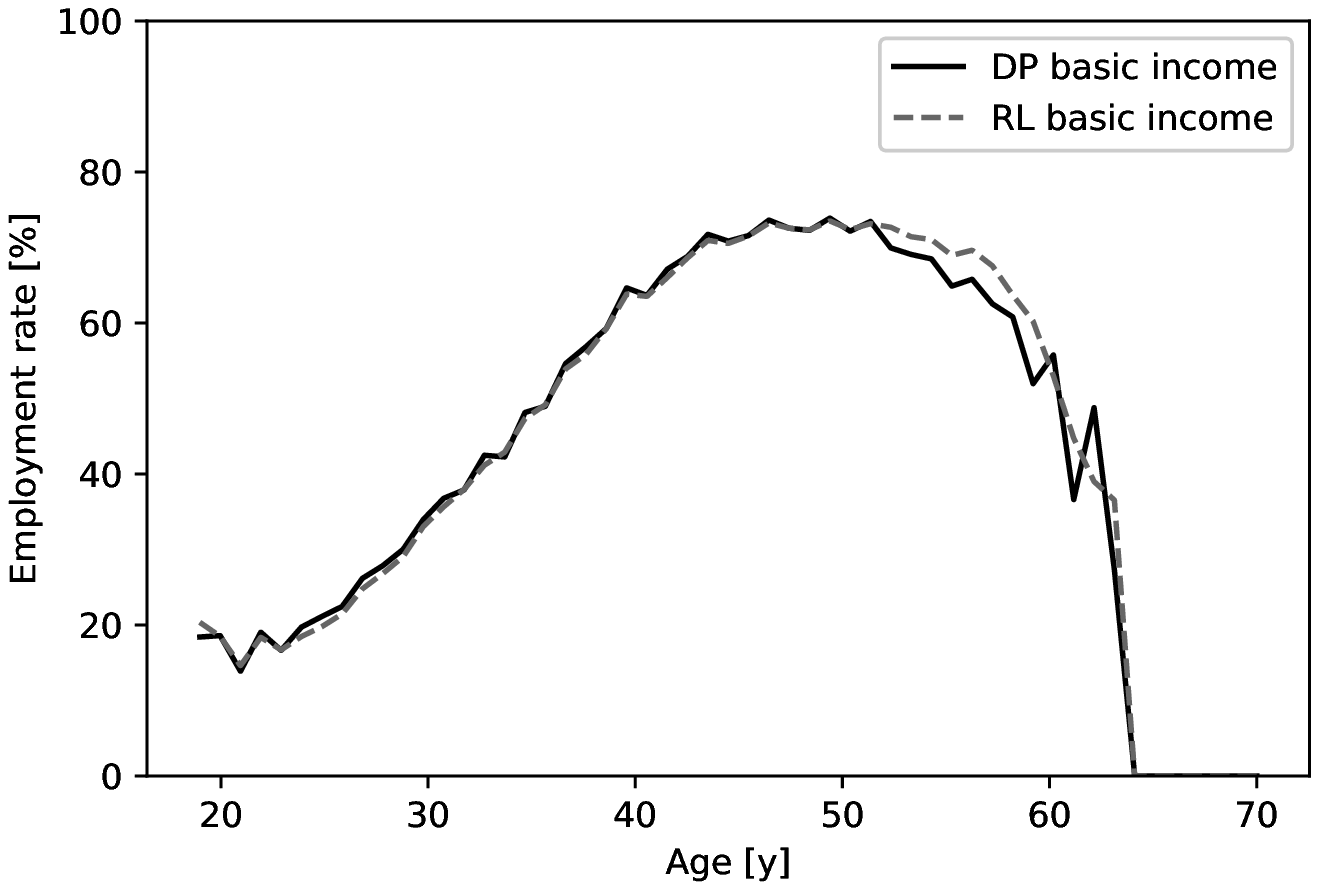}
\caption{Comparison of impacts of the social security reforms on the employment rate analyzed with dynamic programming (DP) and with reinforced learning (RL). (A) Increasing the minimum retirement age to 66 years. (B) Universal basic income. }
\label{fig:fig8}
\end{figure}
 \begin{figure}
\includegraphics[width=7.5cm]{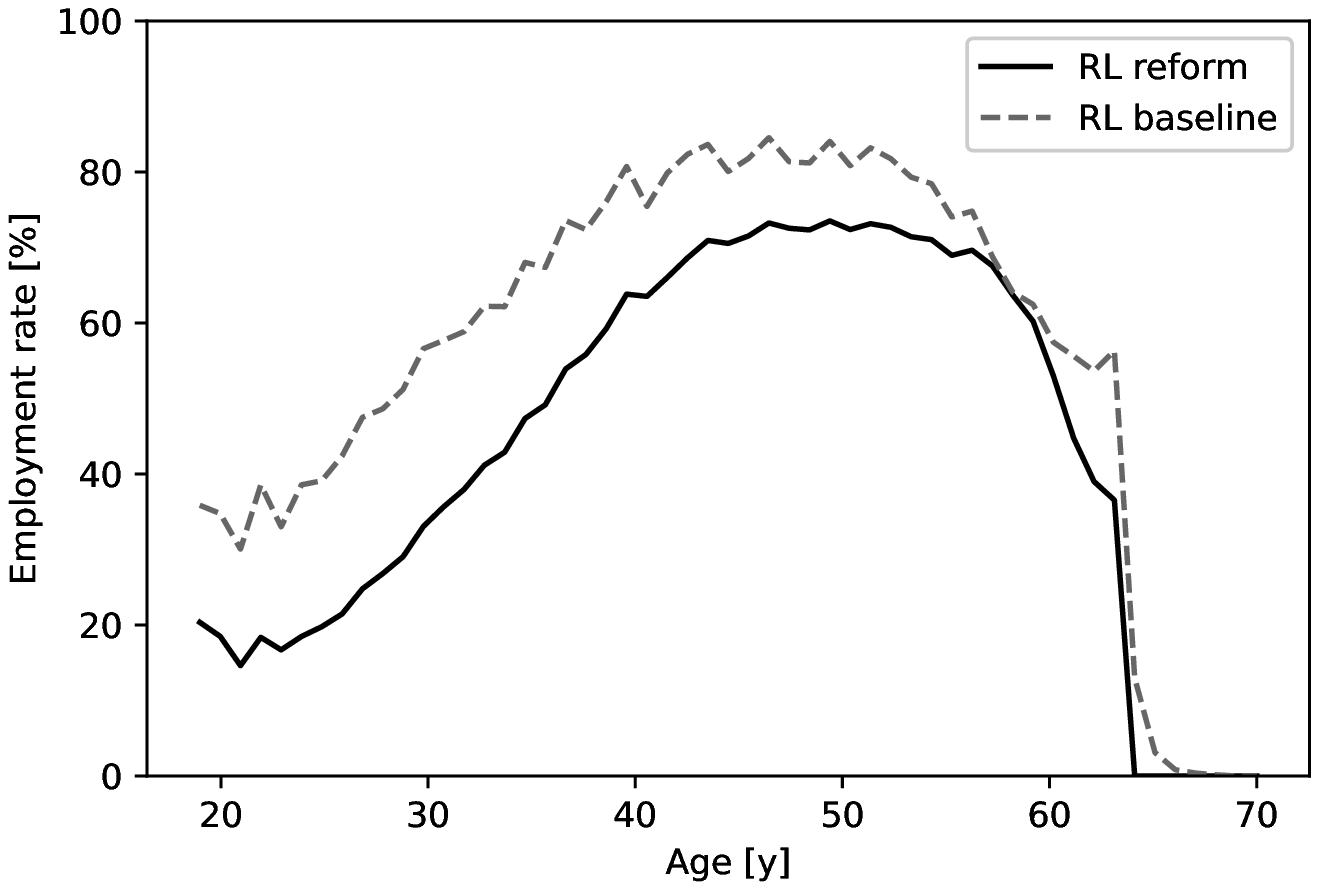}
\includegraphics[width=7.5cm]{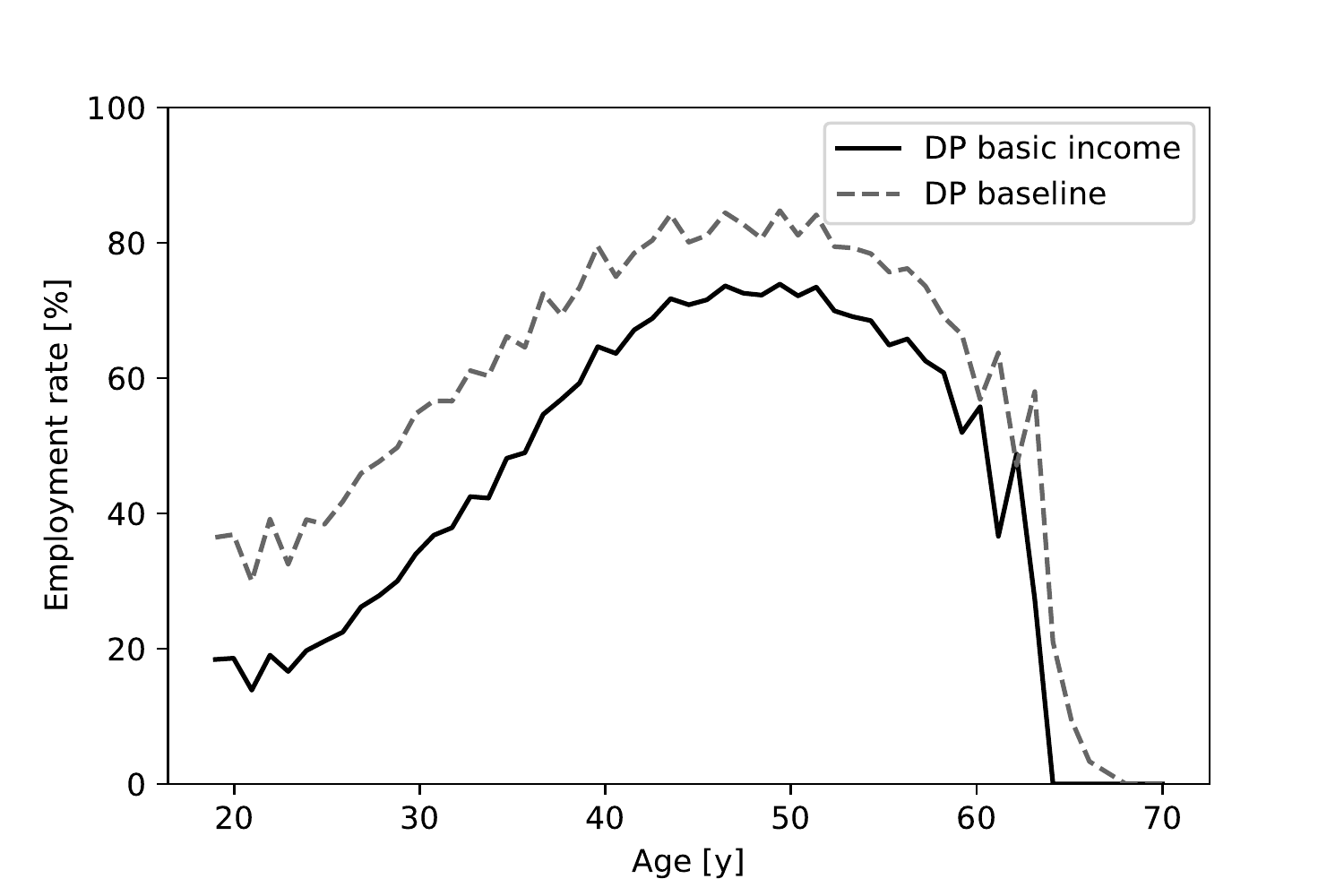}
\caption{Comparison of employment rate in the presence of universal basic income and the baseline in (A) reinforced learning, and (B) in dynamic programming. }
\label{fig:fig7}
\end{figure}
 
The results (Fig. 7 and Table 2) clearly show that UBI, as defined here, would quite significantly reduce the incentives to work and, consequently, reduce employment rate. Similarly, the average discounted utility is significantly reduced, suggesting that UBI would significantly reduce welfare. 
The incentives to work may, in the absence of means-testing, be better in a universal basic income world, however, an increase of taxation reduces the incentive to work. The combined influence of these changes is that employment is significantly reduced (Fig. 7).

Dynamic programming predicts a slightly lower employment rate than reinforced learning from 50 to 60 years of age (Fig. 6) and a slightly lower number of employed persons (Table 2). Similarly to retirement age reform, dynamic programming forecasts a see-saw pattern in employment rate at ages around 60, however, in this case the spikes are quite mild.

Dynamic programming and reinforced learning agree on the impact of UBI reform on employment. DP forecasts that UBI reform would reduce employment by 483,459 person-years, while RL forecasts that UBI reform would reduce employment by 454,817 person-years. 
The difference between forecasts is 28,642 person-years, which is a relatively small difference. 
It is worth stressing that the results presented here are not fully realistic. Still, the sign and possibly also the magnitude
of the impact of the reform on employment would likely stay the same in a more realistic model. 
Preliminary unpublished studies with the life cycle model of Tanskanen (2020a) suggest that this indeed is the case.

\section{Discussion}
\subsection{Comparison of reinforced learning and dynamic programming}
Reinforced learning is a powerful method to approximately solve rich life cycle models of labor supply. The framework presented here enables description of agent behavior in a realistic, detailed model describing social security, incentives to work and taxation. 
For a large life cycle model, such as the one analyzed in Tanskanen (2020a), dynamic programming is not really an option, and reinforced learning is the best available tool for solving the model. This makes it important to understand, how well reinforced learning can reproduce the optimal results.

In a discrete model, DP is known to produce the optimal results. In continuous variable setting, 
the situation is less clear. In the continuum limit, DP converges to the optimal result (e.g., Puterman 1990).
Here, we considered dynamic programming describing behavior in the presence of only a few options for action, but in the presence of
a complex social security scheme, retirement and stochastic wage dynamics. The lattice employed in the discretization was relatively small. 
Nevertheless, DP could solve the model quite well and provided good results. 

The results show that RL can reproduce the results of DP well.
Reinforced learning algorithm ACKTR could approximate the optimal behavior of agents quite well, but not as well as dynamic programming. 
ACKTR produced results comparable to DP in the baseline case and in the two social security reforms. 
When changes in the employment rate due to a social security reform are considered, the direction of the change was correct, but magnitude differ slightly from that found with dynamic programming. This suggests that ACKTR does not always find the global optimum quite as well as dynamic programming. However, in the considered case, the agreement of the methods was sufficient to make the same conclusions on the impact of the reforms using either method.

Our results suggest that reinforced learning produces quite good approximate results to solving a life cycle model. On the other hand, RL enables solving rich life cycle models. Dynamic programming provides a solution that, while approximate, is closer to the true global optimum. The price of this is the computational demands, which only enables solving relatively small life cycle models. 
Hence, one encounters a choice: either solve a more complete model approximately or solve a less complex model more exactly. 
Social security reform are in part needed, because there is a need to simplify the rather complex current social security schemes. 
Hence, approximate methods likely provide the needed tool to analyze the impacts of such reforms.

\subsection{Advantages of reinforced learning}
Reinforced learning algorithms do not need to be told the full description of dynamics of the life cycle model like dynamic programming does. The algorithms will learn the most important parts of the dynamics by sampling the life trajectories. This gives flexibility and ease to developing the life cycle model, since the optimization algorithm does not depend on the particulars of the life cycle model. The use of sampling also reduces the curse of dimensions.

ACKTR is quite robust against anomalies in the code. During the development, ACKTR often produced better results than DP, which should not occur. 
A typical issue was that there was a feature in the life cycle model that was not captured in detail in DP, e.g., an incorrect state transition under certain circumstance. This resulted in suboptimal behavior in the DP model. In the RL model, however, such issues were significantly less important. This is likely  a result of model-independent nature of RL, which makes the use of RL algorithms easier than DP algorithms in practice.
RL gives the modeler a lot more freedom in creating life cycle models than dynamic programming, since RL algorithms are off-the-shelf algorithms that are readily available as high quality implementations that will directly work with the life cycle model as long as the life cycle model interface is written according to a standard framework such as OpenAI Gym (Brockman~\etal 2016).

Here we have shown that ACKTR can be almost as good as dynamic programming in a minimalistic life cycle model. This gives us confidence that this result will apply to more realistic life cycle models. This would be of importance, since reinforced learning algorithms are likely the best currently available tool for solving complex life cycle models.

A large number of reinforced learning algorithms have been proposed, and it is not clear which of the algorithms is the best for solving life cycle models.
In addition to ACKTR, we analyzed several other algorithms, including ACER (Weisz\etal 2018) and A3C (Mnih~\etal 2016). Among these options, ACKTR produced the best results. Reinforced learning algorithms have rapidly been improved in recent years and their convergence properties are understood better (e.g., Wang\etal 2019). The results suggest that the current reinforced learning algorithms can produce good approximations to the optimal results in life cycle models, but may not always do that. Still, more work on the algorithms would be desirable to improve the convergence of the reinforced learning algorithms and understanding their behavior.

\subsection{Future paths and conclusions}

One of the main interests in developing reinforced learning methods for life cycle models is to analyze complex models describing the social security schemes in detail. An interesting open question is which kind of social security schemes are optimal, when behavioral changes and the requirement that the reform is self-financing are taken into account.
The constraints of financing social security require that an optimal social security scheme maximizing welfare should also produce high employment. 
This suggests that UBI considered here is not an optimal social security scheme.

Life cycle models commonly include savings (e.g., Modigliani and Brumberg 1954). This version of the life cycle model does not incorporate saving, however, it would be relatively simple to add to some form of saving in the model. This is likely one of the future development paths.

In conclusion, there is still some way for reinforced learning algorithms to go, but even in their current state they are a valuable tool for solving rich life cycle models. This enables detailed analysis of the impact of the considered social security reform in a realistic setting.

\section*{Acknowledgements}

I thank Niku Määttänen and the two anonymous reviewers for comments and suggestions that greatly improved the manuscript.

\section*{References}

\begin{description}
\item[] Barrett, J. P. (1974) The coefficient of determination—some limitations. The American Statistician, 28(1), 19-20.
\item[] Blau, D.M. (2008) "Retirement and consumption in a life cycle model." Journal of Labor Economics 26: 35-71.
\item[] Brockman, G., Cheung, V., Pettersson, L., Schneider, J., Schulman, J. and Zaremba, W. (2016), “OpenAI Gym”, arXiv:1606.01540.
\item[] Cooley, T.F., and Soares, J. "A positive theory of social security based on reputation." Journal of political Economy 107.1 (1999): 135-160.
\item[] De Wispelaere J., Halmetoja A., Pulkka V.V. "The Finnish Basic Income Experiment: A Primer." The Palgrave International Handbook of Basic Income. Palgrave Macmillan, Cham, 2019. 389-406.
\item[] Fan, J., Wang, Z., Xie, Y. and Zhuoran, X. (2020), “A Theoretical Analysis of Deep Q-Learning”, arXiv:1901.00137
\item[] Hakola, T., Määttänen, N. (2007) ”Vuoden 2005 eläkeuudistuksen vaikutus eläkkeelle siirtymiseen ja eläkkeisiin: arviointia stokastisella elinkaarimallilla.” Eläketurvakeskus.
\item[] Heer, B., Maussner, A. Dynamic General Equilibrium Modeling: Computational Methods and Applications, Springer, 2009.
\item[] Hill, A., Raffin, A., Ernestus, M., Gleave, A., Kanervisto, A., Traore, R., Dhariwal, P., Hesse, C., Klimov, O., Nichol, A., Plappert, M., Radford, A., Schulman, J., Sidor, S. and Wu, Y. (2018), “Stable Baselines”, Github repository, https://github.com/hill-a/stable-baselines
\item[] Jiménez‐Martín, S., Sánchez Martín, A. R. (2007). "An evaluation of the life cycle effects of minimum pensions on retirement behavior", Journal of Applied Econometrics, 22(5), 923-950.
\item[] Kangas, O., Jauhiainen, S., Simanainen, M., Ylikännö, M. (2019). The basic income experiment 2017–2018 in Finland: Preliminary results.
\item[] Kotamäki, M., Mattila, J, Tervola, J. (2018) “Distributional Impacts of Behavioral Effects – Ex-Ante Evaluation of the 2017 Unemployment Insurance Reform in Finland“, Microsimulation 11(2) 146-168
\item[] Mnih, V., Kavukcuoglu, K., Silver, D., Rusu A., Veness J., Bellemare M.G., Graves A., Riedmiller M., Fidjeland A.K., Ostrovski G., Petersen S., Beattie C., Sadik A., Antonoglou I., King H., Kumaran D., Wierstra D., Legg S. and Hassabis D.  (2015) Human-level control through deep reinforcement learning. Nature 518, 529–533. https://doi.org/10.1038/nature14236
\item[] Mnih, V., Badia, A.P., Mirza, M., Graves, A., Lillicrap, T., Harley, T., Silver, D. Kavukcuoglu, K.. (2016). Asynchronous Methods for Deep Reinforcement Learning. Proceedings of The 33rd International Conference on Machine Learning, in Proceedings of Machine Learning Research 48:1928-1937 Available from http://proceedings.mlr.press/v48/mniha16.html .
\item[] Modigliani, F., and Brumberg, R. (1954). Utility analysis and the consumption function: An interpretation of cross-section data. Franco Modigliani, 1(1), 388-436.
\item[] Määttänen, N. (2014), “Evaluation of pension policy reforms based on a stochastic life cycle model”, in book Lassila, Määttänen and Valkonen (2014): Linking retirement age to life expectancy – what happens to working lives and income distribution?
\item[] Pilipiec, P., Groot, W., Pavlova, M. (2020). The Effect of an Increase of the Retirement Age on the Health, Well-Being, and Labor Force Participation of Older Workers: a Systematic Literature Review. Journal of Population Ageing, 1-45.
\item[] Puterman, M.L. (1990) "Markov decision processes." Handbooks in operations research and management science 2: 331-434.
\item[] Rust, J. P. (1989), “A dynamic programming model of retirement behavior”, The economics of aging. University of Chicago Press, 359–404.
\item[] Rust, J. (1997). Using randomization to break the curse of dimensionality. Econometrica: Journal of the Econometric Society, 487-516.
\item[] Sutton, R.S., and Barto, A.G., Reinforcement learning: An introduction, MIT press, 2018
\item[] Tanskanen, A.J. (2020a) Ty{\"o}llisyysvaikutuksien arviointia teko{\"a}lyll{\"a}: Unelmoivatko robotit ansiosidonnaisesta sosiaaliturvasta?, Kansantaloudellinen Aikakauskirja 2, 292-321. In Finnish.
\item[] Tanskanen, A.J. (2020b) Kommentti Viherkentälle eläkemaksujen veroluonteesta, Kansantaloudellinen Aikakauskirja 4, 637-641. In Finnish.
\item[] Tanskanen, A. J. (2019a), life cycle model, Github repository, https://github.com/ajtanskanen/lifecycle-rl (Accessed 2020-Oct-23)
\item[] Tanskanen, A. J. (2019b), Finnish social security and taxation as a Gym-environment, Github repository, https://github.com/ajtanskanen/econogym (Accessed 2020-Oct-23)
\item[] Tanskanen, A. J. (2019c), Benefits - Python module that makes analysis of social security easy, Github repository, https://github.com/ajtanskanen/benefits (Accessed 2020-Oct-23)
\item[] Official Statistics of Finland (OSF) (2020): Population structure. ISSN=1797-5395. Helsinki: Statistics Finland [referred: 2020-oct-2]. Access method: http://www.stat.fi/til/vaerak/index\_en.html
\item[] Wang, L., Cai, Q., Yang, Z. and Wang, Z. (2019), “Neural policy gradient methods: Global optimality and rates of convergence”, arXiv preprint arXiv:1909.01150.
\item[] Yang, Z., Chen, Y., Hong, M. and Wang, Z. (2019), “On the Global Convergence of Actor-Critic: A Case for Linear Quadratic Regulator with Ergodic Cost”, arXiv:1907.06246
\item[] Weisz, G., Budzianowski, P., Su, P. H., Gašić, M. (2018). Sample efficient deep reinforcement learning for dialogue systems with large action spaces. IEEE/ACM Transactions on Audio, Speech, and Language Processing, 26(11), 2083-2097.
\item[] Wu, Y., Mansimov, E., Grosse, R. B., Liao, S. and Ba, J. (2017), “Scalable trust-region method for deep reinforcement learning using kronecker-factored approximation”, Advances in neural information processing systems: 5279–5288.
\end{description}

%
%

\end{document}